\documentclass[a4paper]{iopart}
\pdfoutput=1
\usepackage{amsmath}
\usepackage{hyperref}
\usepackage{amsmath,amsthm, amssymb}
\usepackage{amsfonts}
\usepackage{pgf}
\usepackage{slashed}
\usepackage{graphicx}
\usepackage{dcolumn}
\usepackage{bm}
\newcommand{\ket}[1]{\left|#1\right\rangle}

\newcommand{\mean}[1]{\left\langle #1\right\rangle}

\newcommand{\kk}{K/\kbar}
\newcommand{\jj}{J/\kbar}
\usepackage{color}
\newcommand{\kbar}{\mathchar'26\mkern-9mu k}
\newcommand{\nep}{\textrm{e}}
\newcommand{\ud}{\mathrm{d}}

\definecolor{caribbeangreen}{rgb}{0.0, 0.8, 0.6}
\newcommand\sn[1]{{#1}}

\begin{document}


\title{{Slow heating in a quantum coupled kicked rotors system}}
%
\author{Simone Notarnicola}
\address{Dipartimento di Fisica ed Astronomia ``Galileo Galilei'' \& INFN, via Marzolo 8, I-35131 Padova, Italy.}
%
%
\author{Alessandro Silva}
\address{%
 SISSA, Via Bonomea 265, I-34136 Trieste, Italy
}%

\author{Rosario Fazio }
\address{ Abdus Salam ICTP, Strada Costiera 11, I-34151 Trieste, Italy }
\address{Dipartimento di Fisica, Universit\`a di Napoli ``Federico II", Monte S. Angelo, I-80126 Napoli, Italy}
%
%
\author{Angelo Russomanno }%
\address{ Abdus Salam ICTP, Strada Costiera 11, I-34151 Trieste, Italy }
\address{ Max-Planck-Institut f\"ur Physik Komplexer Systeme, N\"othnitzer Strasse 38, D-01187, Dresden, Germany }

\date{\today}

\begin{abstract}
We consider a finite-size periodically driven quantum system of coupled kicked rotors which exhibits two distinct regimes in parameter space: a dynamically-localized one with kinetic-energy saturation in time and a chaotic one with unbounded energy absorption (dynamical delocalization). We provide numerical evidence that the kinetic energy grows subdiffusively in time in a parameter region close to the boundary of the chaotic dynamically-delocalized regime. We map the different regimes of the model via a spectral analysis of the Floquet operator and investigate the properties of the Floquet states in the subdiffusive regime. We observe an anomalous scaling of the average inverse participation ratio (IPR) analogous to the one observed at the critical point of the Anderson transition in a disordered system. We interpret the behavior of the IPR and the behavior of the asymptotic-time energy as a mark of the breaking of the eigenstate thermalization in the subdiffusive regime. Then we study the distribution of the kinetic-energy-operator off-diagonal matrix elements. We find that in presence of energy subdiffusion they are not Gaussian and we propose an anomalous random matrix model to describe them.
\end{abstract}

\pacs{Valid PACS appear here}
\maketitle

\section{Introduction}

Chaos and energy absorption are intimately related. Efficient energy absorption occurs when the driving is resonant with some natural frequencies of the system and chaos develops around resonances~\cite{Berry_regirr78:proceeding,Lichtenberg}. If the driving amplitude is large enough, all the phase space is chaotic and the energy diffusively increases up to the {so-called} infinite-temperature value. For smaller values of the amplitude, the KAM theorem states that only a part of the phase space is chaotic~\cite{salamon2004kolmogorov,Arnold-Avez:book}. Nevertheless, in the many-body case, this fact gives rise to a chaotic web uniformly spread in the phase space. Along this web, diffusion in phase space can occur giving rise to a slow energy increase up to the $T=\infty$ value, possibly after a prethermal behavior~\cite{Lichtenberg,konishi,chiri_vov,Ata_ema,Oven}.

In the quantum case, the route leading to chaos, energy absorption and $T=\infty$ thermalization is different. In this case a central role is played by the properties of the Floquet states, the eigenstates of the stroboscopic {periodically-driven} dynamics~\cite{Shirley_PR65,Samba}. There is thermalization when these states are strongly entangled and locally equivalent to the thermal $T=\infty$ density matrix. In this case the local observables asymptotically relax to the $T=\infty$ thermal ensemble with fluctuations vanishing in the thermodynamic limit~\cite{Rigol_PRX14,Ponte_AP15,srednicki_jpa99,Polkovnikov_quergo:booklet,Russomanno_EPL}. This is a form of the so-called eigenstate thermalization~\cite{Sred_PRE94,Deutsch_PRA91,Rigol_Nat}, where thermal behavior appears because of the properties of the eigenstates of the dynamics. There is a chaotic and thermalizing behavior when the eigenstates of the dynamics behave as the eigenstates of a random matrix~\footnote{This is rigorously true in the case of a driven system. In the autonomous case, the eigenstates of the Hamiltonian behave as the eigenstates of a random \textit{banded} matrix~\cite{srednicki_jpa99}, in order to ensure energy conservation.}~\cite{Bohigas_PRL84,Berry_LH84,Haake:book}.


Usually, there is correspondence between classical and quantum chaos. If the classical dynamics is chaotic and thermalizing, the quantum eigenstates obey eigenstate thermalization and the Hamiltonian (or the Floquet Hamiltonian in the driven case -- see later) behaves as a random matrix in physical bases~\cite{CHIRIKOV198877,Haake:book,Berry77,Pechukas,feingold_PRL,prosen_AP94,PhysRevE.50.888,Bohigas_PRL84,Eckardt_PRE95,sred_therm}.
Despite these expectations, there are exceptions. The best known is the quantum kicked rotor~\cite{Chirikov1979263, chirikov,PhysRevLett.49.509, PhysRevA.29.1639}. This non-integrable model describes a free rotor perturbed by a time-periodic  kick with strength $K$. In the classical case, when $K$ is below some threshold, the system is quasi-integrable according to the KAM theory; here there is no energy absorption. On the opposite, when $K$ is very large, the phase space is fully chaotic and the system absorbs energy which increases linearly and diffusively in time. The quantum behavior is completely different: quantum interference hinders energy absorption for all values of $K$ and the kinetic energy linearly increases until a saturation value is reached~\cite{chirikov_argument}. This phenomenon is called dynamical localization and the reason behind it is that the wave function of the rotor is localized in the angular momentum representation. This form of localization has been found~\cite{PhysRevLett.49.509} to be equivalent to the Anderson localization of a particle hopping in a static, one dimensional disordered lattice~\cite{PhysRev.109.1492}.

This is a remarkable result and many efforts have been devoted to see if dynamical localization survives when many interacting rotors are considered. When many interacting classical rotors are considered, chaos has a stronger effect on energy absorption. As stated above, even a partly chaotic phase space is enough to have Arnold diffusion, and this fact translates in a diffusive long-term dynamics, possibly after a transient~\cite{Nekhoroshev1971,konishi, PhysRevA.40.6130, 2018arXiv180101142R,Ata_ema}. In the quantum case the behavior is different and dynamical localization can persist for a finite number of rotors~\cite{nostrolavoro,2rot_1,2rot_2, 2rot_3, Adachi_PRL88}. Nevertheless, it disappears in the thermodynamic limit, when the number of rotors tends to infinity~\cite{nostrolavoro} (although it can persist in the thermodynamic limit in other systems~\cite{rozenbaum_e,rylands}). 

Even in absence of localization, quantum effects can alter the energy absorption dynamics. A way in which this can happen is the induction of a sub-diffusive energy increase, a regime where the  kinetic energy does not increase linearly and diffusively but as a power law with exponent smaller than 1. This regime up to now has only been found in  mean-field studies~\cite{nostrolavoro} and one might be tempted to think that it can appear only in the peculiar conditions where mean field is valid (thermodynamic limit, infinite-range interactions).

A hint that this might not be true comes from studies of localization in space. In Anderson models with mean-field interactions a breaking Anderson localization and consequent subdiffusion in space have been \sn{observed~\cite{ shepelyanski_prl93, shepelyanski_prl08,Laptyeva_2014,PhysRevB.89.060301} even away from the mean-field limit}. Indeed, subdiffusion in space for ergodic disordered models near the MBL transition has been theoretically predicted in~\cite{marko} and experimentally observed in~\cite{Matthew}.

 \begin{figure}
    \begin{center}
     \vspace{10pt}
    \includegraphics[width=8cm]{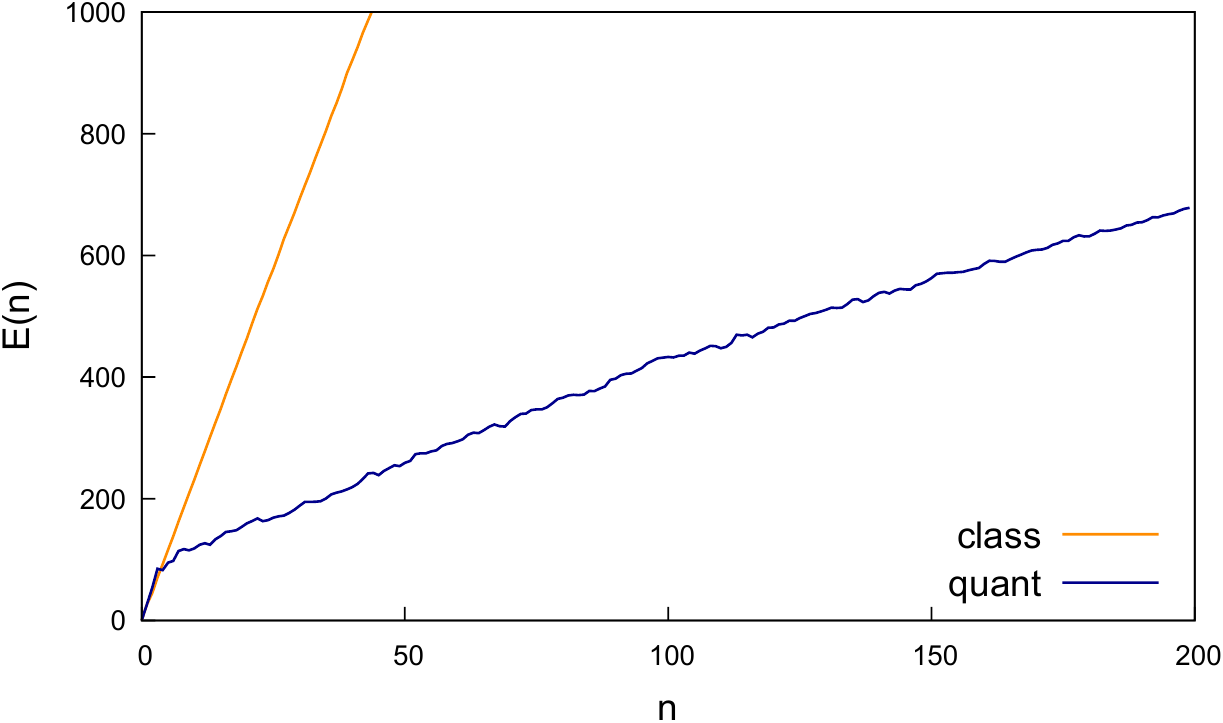}
   \end{center}
 \caption{\label{fig:class_quant_dyn} Growth of the kinetic energy per rotor $E(n)$: it is linear in the classical system (light-colored line) and subdiffusive in the quantum one (dark-colored line). Other parameters: $\kk=2.0,\, \jj=0.3$, $L=3$, $\kbar=5.0$.}
  \end{figure} 
  
\sn{ 
Motivated by the emergence of subdiffusive behaviors in disordered systems, in the boundary region between localized and ergodic regimes, we want to study the case of a periodically-driven system exhibiting two separated regimes, a dynamically-localized one and an ergodic one.
We focus on the case of $L=3$ interacting quantum kicked rotors: indeed, this is the minimal case in which a transition from a dynamically-localized regime to a delocalized one can be observed \cite{nostrolavoro}.
As it appears in other periodically-driven  \cite{Qin2017,refId0,0295-5075-96-3-30004} as well as time-independent, few-body models \cite{Yusipov2017}, we show that in our system subdiffusion occurs not in the space  but in the momentum domain, manifesting itself in the energy absorption process. In particular, we observe a parameter region where quantum effects determine a dynamics different from the classical one even in the absence of localization: indeed, a subdiffusive power-law heating is observed in contrast with the linear, diffusive classical one.
 We show that the subdiffusion manifests in a region of the parameters space where the system is delocalized situated close to the boundary with the localized region, in analogy with the results relative to systems undergoing the MBL transition.}

An example of this subdiffusion is shown in Fig.~\ref{fig:class_quant_dyn} where we compare it with the corresponding classical dynamics. We see that after a transient, the quantum and the classical evolutions are completely different,  the first being  subdiffusive and the second  diffusive (as appropriate for an essentially chaotic classical dynamics). Subdiffusion is a genuinely quantum phenomenon originating from interference, with no classical counterpart.
It is therefore important to connect this subdiffusion to the quantum properties of the model, especially to its Floquet states which are the eigenstates of the stroboscopic dynamics.

We find that, in the subdiffusive regime, the kinetic-energy matrix elements in the Floquet basis show anomalous distributions, different from the pure Gaussians~\cite{Berry77,Pechukas,feingold_PRL,prosen_AP94,PhysRevE.50.888,Bohigas_PRL84,Eckardt_PRE95,sred_therm} of the fully chaotic and thermalizing case. This behavior has already been observed in ergodic disordered systems, near the transition to many-body localization, where two-time correlators show a subdiffusive behavior in space~\cite{Luitz_PRL16,Roy_PRB18}. We can provide an interpretation of the anomalous distribution of the matrix elements. Indeed we see the same distributions if we consider a random-matrix model with fluctuations of the matrix elements decaying as a power law with the distance from the diagonal. 

\begin{figure}
  \begin{center}
    \includegraphics[width=8cm]{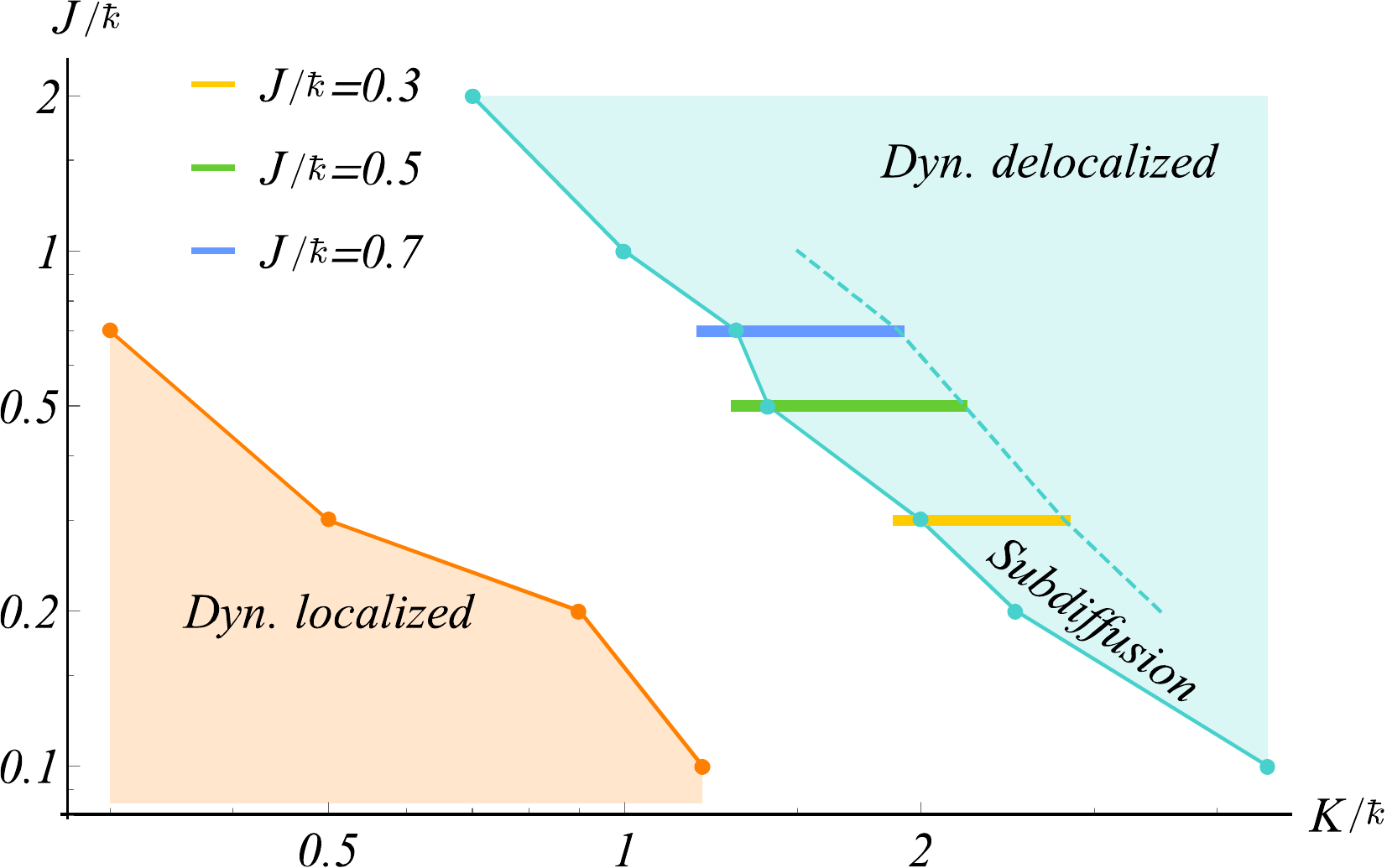}
  \end{center}
  \caption{\label{fig:phase_diagram} Dynamical regimes of the model. {$K$ and $J$ are the kicking strengths; $K$ is the amplitude of a kick which acts on each rotor separately while $J$ couples nearest-neighboring rotors. $\kbar$ is the effective Planck constant. All these definitions are in Eq.~\eqref{eq:def_ham} and in the discussion below. The dynamically localized 
  and
  delocalized regimes coincide with the regions in which the averaged level spacing ratio $\mean r$ assumes the Poisson value or  the Circular Orthogonal Ensemble one}. The numerics suggests a limited growth of the kinetic energy and an unbounded heating in the dynamically-localized {regime} and in the delocalized 
  {one}
  respectively. 
  The horizontal lines represent the intervals in $\kk$ where the subdiffusion of the kinetic energy has been observed for different values of $\jj$.  System size $L=3$, effective Planck's constant $\kbar=5.0$.}
\end{figure}

The paper is organized as follows. In Section~\ref{modello:sec} we describe the model and the details of the numerical exact-diagonalization analysis we perform on it. In Section~\ref{risultati:sec} we show the power-law behavior in time of the kinetic energy. We map 
{the different regimes we observe}
in 
Fig.~\ref{fig:phase_diagram}: We find a dynamically-localized {regime} and a delocalized one where numerics suggests unbounded energy absorption. {Subdiffusion occurs close to the boundary of the dynamically-delocalized region in parameter space}. We focus on the subdiffusive regime  and show that in this regime the eigenstate thermalization is broken. We can see this fact by studying the properties of the Floquet states: They show 
 {long tails in the Inverse Participation Ratio (IPR) distributions and large IPR fluctuations}. The breaking of eigenstate thermalization is reflected also in the asymptotic value of the energy with a truncated Hilbert space.
We also study the
off-diagonal matrix elements of the kinetic-energy operator in the Floquet basis: We find that the distributions of these matrix elements have long tails and are different from a Gaussian, marking the fact that the dynamics is not perfectly chaotic (similar distributions appear in cases of anomalous thermalization in many-body systems~\cite{Luitz_PRL16,Roy_PRB18}).  In Section~\ref{randommat:sec} we interpret the distributions of the off-diagonal matrix elements of the kinetic-energy operator: We show that they can be derived from a model based on a random matrix with fluctuations of the matrix elements decaying as a power law with the distance from the diagonal.  In Section~\ref{conc:sec} we draw our conclusions.

\section{The model} \label{modello:sec}
The model we consider is described by the following Hamiltonian:
\begin{align}\label{eq:def_ham}
&\hat H(t) = \hat H_0+\sum_{n=-\infty}^{+\infty}\delta(t-n T ) \hat V({\boldsymbol \theta}) \\ \nonumber
&\hat H_0 = \sum_{j=1}^L\frac{\hat p_j^2}{2 I_{\rm m.~i.}}\,,\,\,
\hat V({\boldsymbol \theta}) = \sum_{j=1}^L\left[K \cos \hat \theta_j+J\cos(\hat\theta_j-\hat\theta_{j+1})\right]\,,
\end{align}
with a time-independent kinetic term and an angle-dependent periodic perturbation; in particular, $K$ is the strength of the kick acting on each single rotor and $J$ is the amplitude of a nearest-neighbor coupling term.
With a proper redefinition \cite{Haake:book} of the operators and the constants $K$ and $J$ in the Hamiltonian in Eq.~(\ref{eq:def_ham}) we set the time period $T$ and the inertia momentum $I_{\rm m.~i.}$ to 1.
We choose periodic boundary conditions such that $L+1\equiv 1$;
we impose the canonical commutation relations to the angle and angular momentum operators, namely $[\hat p_j, \hat p_k]=[\hat\theta_j, \hat \theta_k]=0$ and $[\hat\theta_j, \hat p_k]=i\kbar\,\delta_{j\,k}$, where $\kbar=\hbar\, T/I_{\rm m.~i.}$. 
 The Hilbert space in which the system evolves is 
$\mathcal{H}=\bigotimes_{j=1}^L \mathcal{H}_j$ where $\mathcal{H}_j$ is the Hilbert space of a single rotor. The basis vectors in the angular momentum representation are indicated with  $|{\bf p}\rangle=|p_1,\,...\,p_L\rangle$, where $\hat p_j$ is the angular momentum of the $j$-th rotor. The spectrum of the operators $\hat p_j$ is unbounded; moreover, due to the $2 \pi$-periodicity of the wave function with respect to each of the $\hat \theta_j$ operators, it is discrete with eigenvalues $p_j=\kbar m_j$ with $m_j\in \mathbb{Z}$.

In general, the stroboscopic dynamics of periodically-driven quantum systems is studied by introducing the Floquet operator $\hat U_F$, defined as the time propagator over one period. In our case, we have
\begin{equation}\label{eq:simm_Uf}
\hat U_F =
\nep^{-i\hat{H}_0/(2\kbar)}\nep^{- i\,V(\hat{\boldsymbol \theta})/\kbar}\nep^{- i\hat{H}_0/(2\kbar)}\,,
\end{equation}
where we set the initial time in the middle of the free evolution in order to make the time inversion symmetry explicit. We can numerically diagonalize this unitary operator in the form
\begin{equation}
  \hat U_F\ket{\psi_\beta}=\nep^{-i\mu_\beta}\ket{\psi_\beta}\,,
\end{equation}
where the eigenstates $|\psi_\beta\rangle$ are the so-called Floquet states and the eigenvalue phases $\mu_\beta$ the corresponding quasienergies~\cite{Shirley_PR65,Samba}. 

Our analysis is based on the full exact diagonalization of the Floquet operator, through which we infer its spectral properties and we compute the exact dynamics of the system. 
Since the local Hilbert space $\mathcal{H}_j$ relative to each site has infinite dimension, a local truncation is necessary in order to write the Floquet operator matrix: we fix a maximum value for the angular momentum on each site $m_{\rm max}\in \mathbb{N}$ and consider only angular momentum eigenstates with eigenvalue $|p_j|\leq m_{\rm max}$ so that the dimension of the whole Hilbert space is $M^L$, with $M=2\,m_{\rm max}+1$. We restrict our analysis to the subspace invariant under the following symmetry transformations: spatial translation ($j'=j+1$), spatial inversion ($j'=L-j$) and global momentum parity ($p'_j=-p_j$ $\forall j$), so that the dimension of the Hilbert subspace we deal with is $D=M^L/(4L)$.

\section{Subdiffusion and breaking of eigenstate thermalization} \label{risultati:sec}
{
In this section we study the subdiffusion behavior of the energy (Sec.~\ref{evoleno:sec}) and relate it with the properties of the Floquet states and of the off-diagonal matrix elements of the kinetic energy operator in the Floquet basis. We discuss the last point in Sec.~\ref{distro:sec}. We see that the distributions are not Gaussian, exactly as occurs in cases of anomalous thermalization in many-body systems~\cite{Luitz_PRL16,Roy_PRB18,foini2019eigenstate}. In Sec.~\ref{levelspacing:sec} we study the properties of the Floquet levels and we see that subdiffusion is associated to an average level spacing ratio very near to the ergodic value. Nevertheless, the Floquet states do not obey the eigenstate thermalization, as we see in Sec.~\ref{IPR:sec} by studying the properties of the IPR distribution of the Floquet states. Moreover, for the local Hilbert space truncations $M$ we can reach, the system does not thermalize to $T=\infty$ when there is subdiffusion. This suggests that in our case subdiffusion appears in association with breaking of eigenstate thermalization. This is different from the known many-body cases, where subdiffusion is associated to thermalization in the thermodynamic limit~\cite{Luitz_PRL16,Roy_PRB18,foini2019eigenstate}. Honestly, due to numerical limitations, we do not know if a thermalization behavior is attained in the limit $M\to\infty$ and if this would imply a breaking of subdiffusion in the long time.}
\subsection{Energy evolution} \label{evoleno:sec}
We start studying the kinetic-energy dynamics, showing examples of power-law increase in time of this quantity.
The stroboscopic dynamics of  a given initial state $|\psi_0\rangle$ is given by
$|\psi(n)\rangle = (\hat U_F)^n|\psi_0\rangle$. The observable on which we focus is the kinetic energy of the system per rotor defined as $ E(n) =\langle \psi(n) | \hat H_0 | \psi(n)  \rangle/L$, that can be re-written as
\begin{equation}\label{eq:kin_en}
E(n)=\sum_{\beta,\gamma}\mathcal{H}^{\,0}_{\beta\,\gamma}\,\mathrm{e}^{-in(\mu_\beta-\mu_\gamma)}\,\psi_\beta\psi_\gamma\,,
\end{equation}
where $\mathcal{H}^{\,0}_{\beta\,\gamma}=\langle\psi_\beta|\hat H_0|\psi_\beta\rangle/L$ and $\psi_\beta=\langle\psi_\beta|\psi(0)\rangle$. 
\begin{figure}
   \begin{center}
     \vspace{10pt}
    \includegraphics[width=8cm]{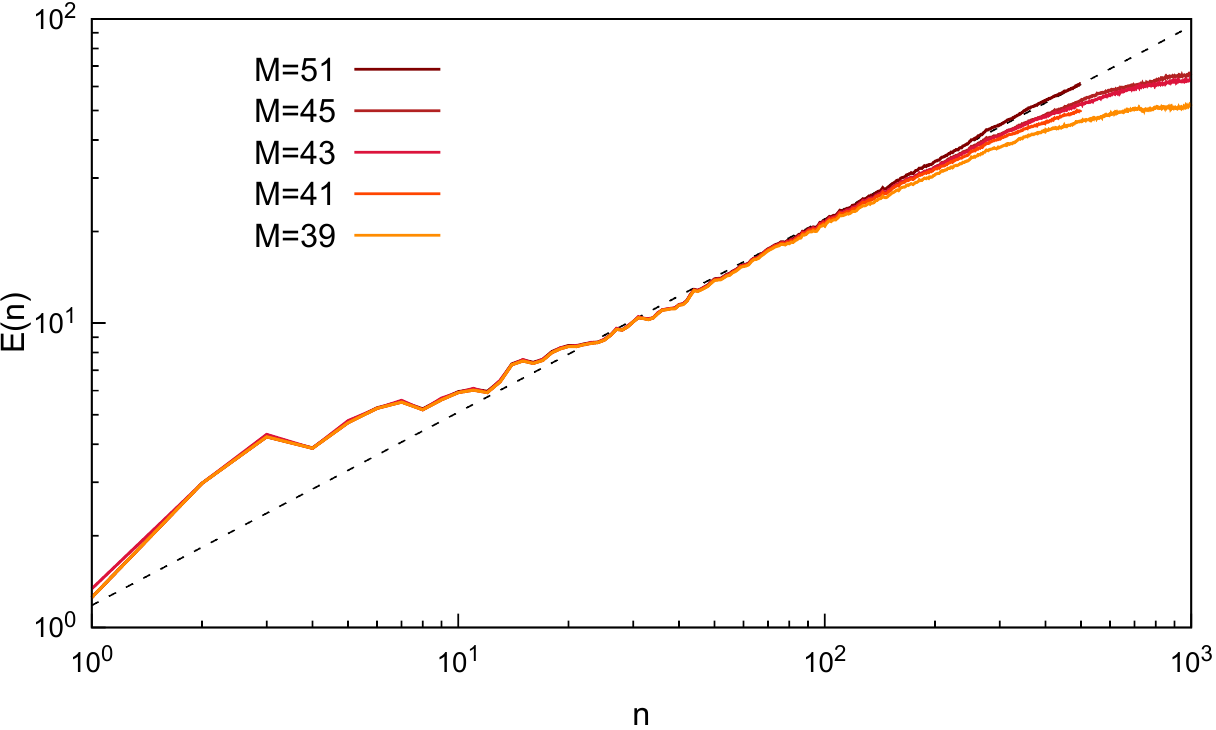}
   \end{center}
  \caption{\label{fig:en_growth} The growth of the kinetic energy is plotted for $\kk=2.2$ and $\jj=0.3$. The power law appears for all the values of the truncation we consider, from $M=39$ (lighter curve) to $M=51$ (darker curve). The time at which the growth stops depends on the saturation value of the energy, which increases with $M$. It is evident that the larger is $M$, the longer the time window where  the power law persists, the better the fit. Other parameters: $L=3$, $\kbar=5.0$.}
\end{figure}
We analyze the dynamics of the system focusing on the kinetic energy operator defined in Eq.~(\ref{eq:kin_en}). We choose as initial state the momentum eigenstate $|\bf{0}\rangle$ in which all the rotors have zero angular momentum. Let us consider Fig.~\ref{fig:en_growth} in which we plot the growth of the kinetic energy for a choice of the values of $\jj$ and $\kk$ and different values of $M$. In this figure we can distinguish two regimes. The first occurs at intermediate times: the dynamics of the system is independent on the truncation, while its duration increases with $M$. The second occurs at long times dynamics: the energy tends to an asymptotic value which depends on the truncation $M$. In the following we separately analyze these two regimes.  { We postpone the analysis of the asymptotic regime to Sec.~\ref{asymptotic:sec} and here we focus on the intermediate-time dynamics.

In order to analyze this regime we measure the power-law exponent of the kinetic energy growth for different values of $\jj$ and $\kk$. We observe a region in the parameter space in which the heating process in the classical system is very different from the quantum one, as in the first case heating is linear, while in the second we find $E(n)\sim n^\alpha$ with $\alpha<1$ (see Fig.~\ref{fig:class_quant_dyn}).  
The power-law heating is not due to the local truncation, as we checked by computing the dynamics for several values of $M$, as it is shown  in Fig.~\ref{fig:en_growth}. {Although we cannot exclude {\it a priori} that different regimes may arise at longer times, the available time scales and the truncation values are enough to claim the existence of a genuine quantum regime, different from the classical one, as it is evident from Fig.~\ref{fig:class_quant_dyn}.}
We repeat the same procedure for a grid of values of $\jj$ and $\kk$ and  we compute the power law coefficient. In Fig.~\ref{fig:pl_exponents} we plot $\alpha$ as a function of $\kk$ for different values of $\jj$; the lower boundary in the interval of $\kk$ values  is that of the dynamically-delocalized regime, while the upper one is determined by numerical limitations. {Even though a clear dependence of $\alpha$ from the parameters $K$ and $J$ is missing we notice that $\alpha$ seems to increase as $J$ is increased.} 
\begin{figure}
\vspace{10pt}
\begin{center}
\includegraphics[width=8cm]{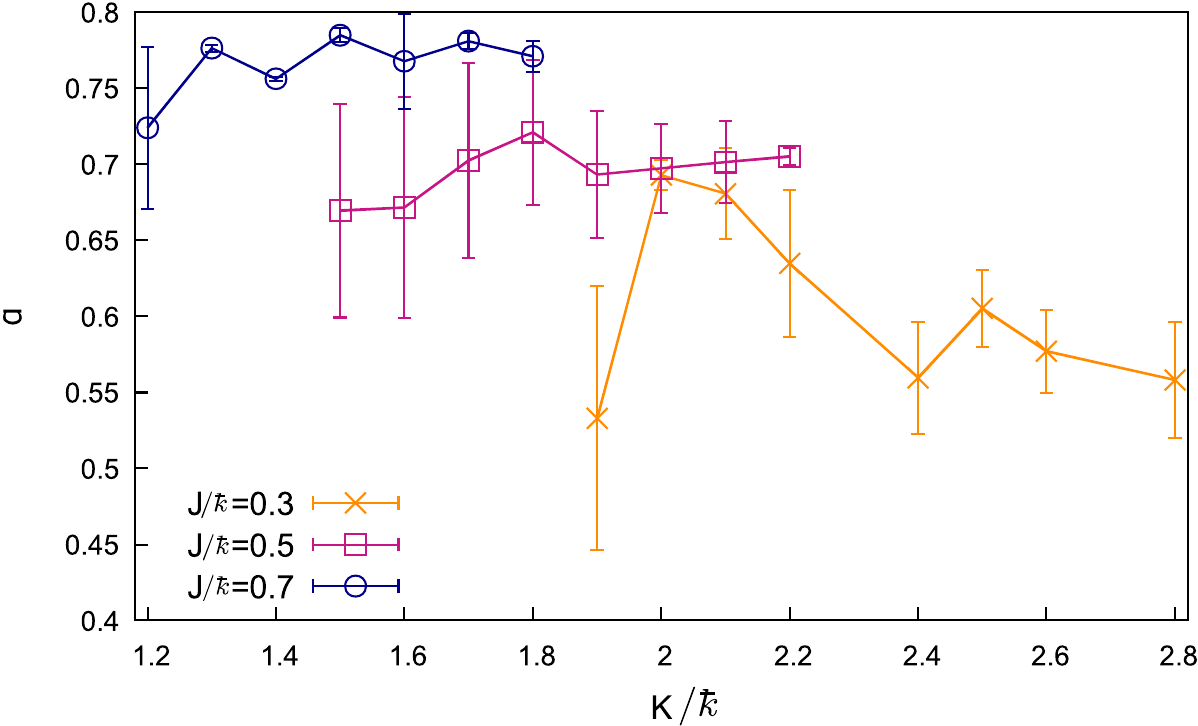}
\end{center}
\caption{\label{fig:pl_exponents} The power-law exponent $\alpha$ is plotted for $\jj=0.3,\,0.5,\,0.7$ and some values of $\kk$. It appears that the exponent increases as $\jj$ is increased while a clear behavior is not clear in the interval of $\kk$ we can consider within the maximum truncation $M$ we can achieve in our simulations. \sn{The error bars are obtained by measuring the exponents over different time intervals whose length is one order of magnitude and then evaluating their semi-dispersion}. Other parameters: $L=3$, $\kbar=5.0$.}
\end{figure}
\subsection{Distribution of the off-diagonal matrix elements}\label{distro:sec}
Subdiffusion corresponds to the breaking of the perfect chaoticity of the dynamics. From one side this can be already seen from the fact that a perfectly chaotic classical dynamics leads to diffusion and not subdiffusion. From the quantum perspective this can be seen by observing the properties of the Floquet states and noticing that {they have properties different from perfect eigenstate thermalization}. In order to do that, we start considering that the diffusion dynamics is given by the off-diagonal matrix elements of the energy operator $\hat{H}_0$ in the Floquet basis (see Eq.~\eqref{eq:kin_en}). This suggests to inquire the behavior of the distribution of the off-diagonal elements $\mathcal{H}^{\,0}_{\beta\gamma}$ with $\gamma\neq\beta$, in order to interpret the power-law increase behavior of the energy (a similar analysis was performed to interpret an anomalous thermalization behavior in~\cite{Luitz_PRL16} possibly associated with subdiffusion behavior in space~\cite{Roy_PRB18}). If we had a perfect chaotic behavior, the operators expressed in the basis of the Floquet states should behave as a perfect random matrix~\cite{Peres_PRA84}, therefore the matrix elements $\mathcal{H}^{\,0}_{\beta\gamma}$ should be distributed according to a Gaussian.

We plot the distribution of $\mathcal{H}^{\,0}_{\beta\gamma}/\Sigma$ (where $\Sigma$ is the variance of the distribution of the $\mathcal{H}^{\,0}_{\beta\gamma}$) for many subdiffusive cases in the upper panels of Fig.~\ref{fig:distrib}. We find indeed a significant deviation from a Gaussian behavior, as it should have been expected being the corresponding behavior of $E(n)$ different from the perfectly chaotic diffusion. 
(One of the distributions of Fig.~\ref{fig:distrib}--right panel corresponds to the subdiffusion depicted in Fig.~\ref{fig:en_growth}.) In order to do a comparison, we plot in the lower panel of Fig.~\ref{fig:distrib} the distribution for a case which is fully thermalizing without any subdiffusion (at least for the truncations we have access to). We see that it is an (almost) perfect Gaussian, {in agreement with the expectations from Random Matrix Theory}. From this comparison we  see that the behavior of the distribution of the off-diagonal elements and the time behavior of the energy are intimately connected. 
%
\begin{figure}
\vspace{10pt}
\begin{center}
\begin{tabular}{cc}
  \hspace{-1cm}\includegraphics[width=8cm]{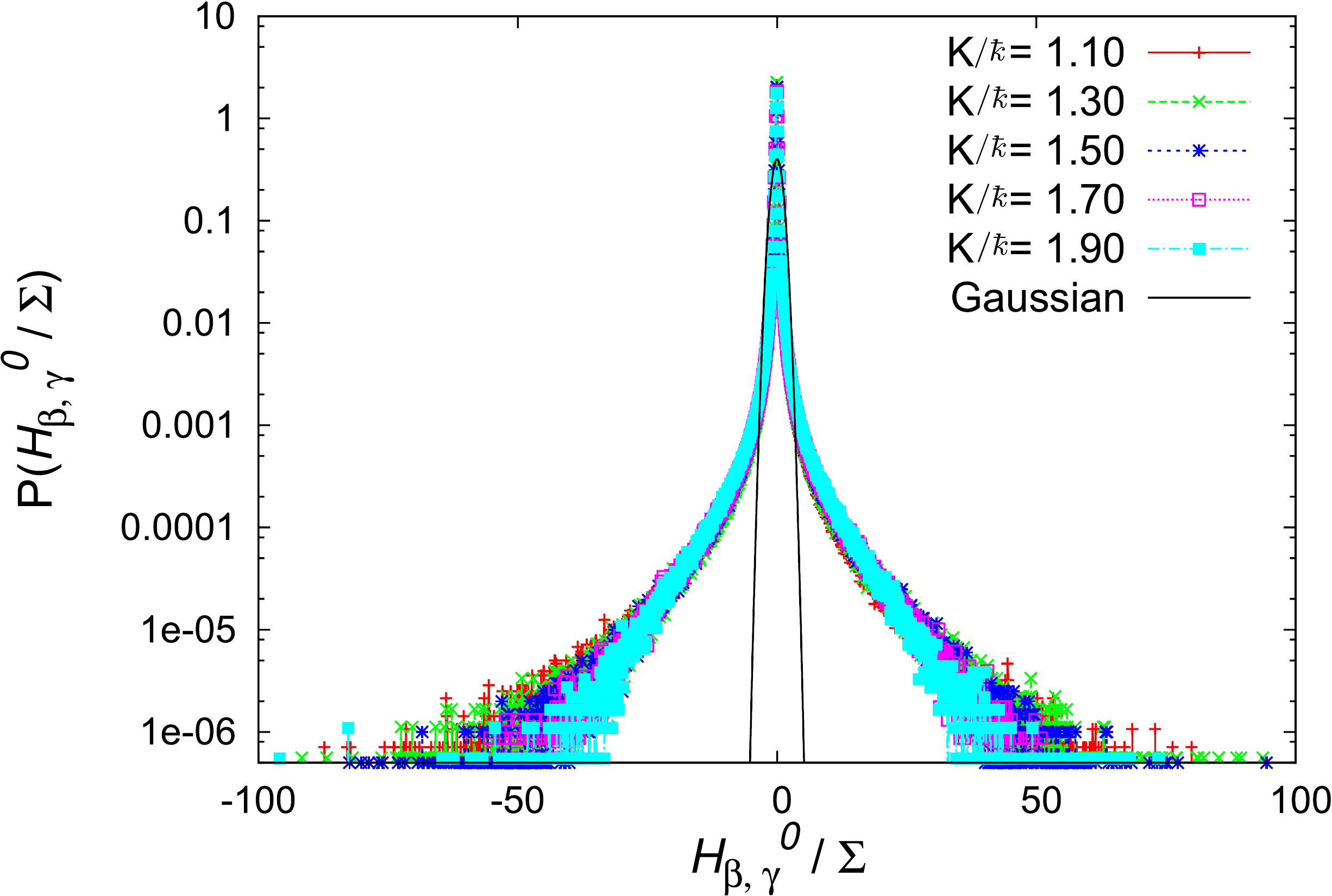}&
  \includegraphics[width=8cm]{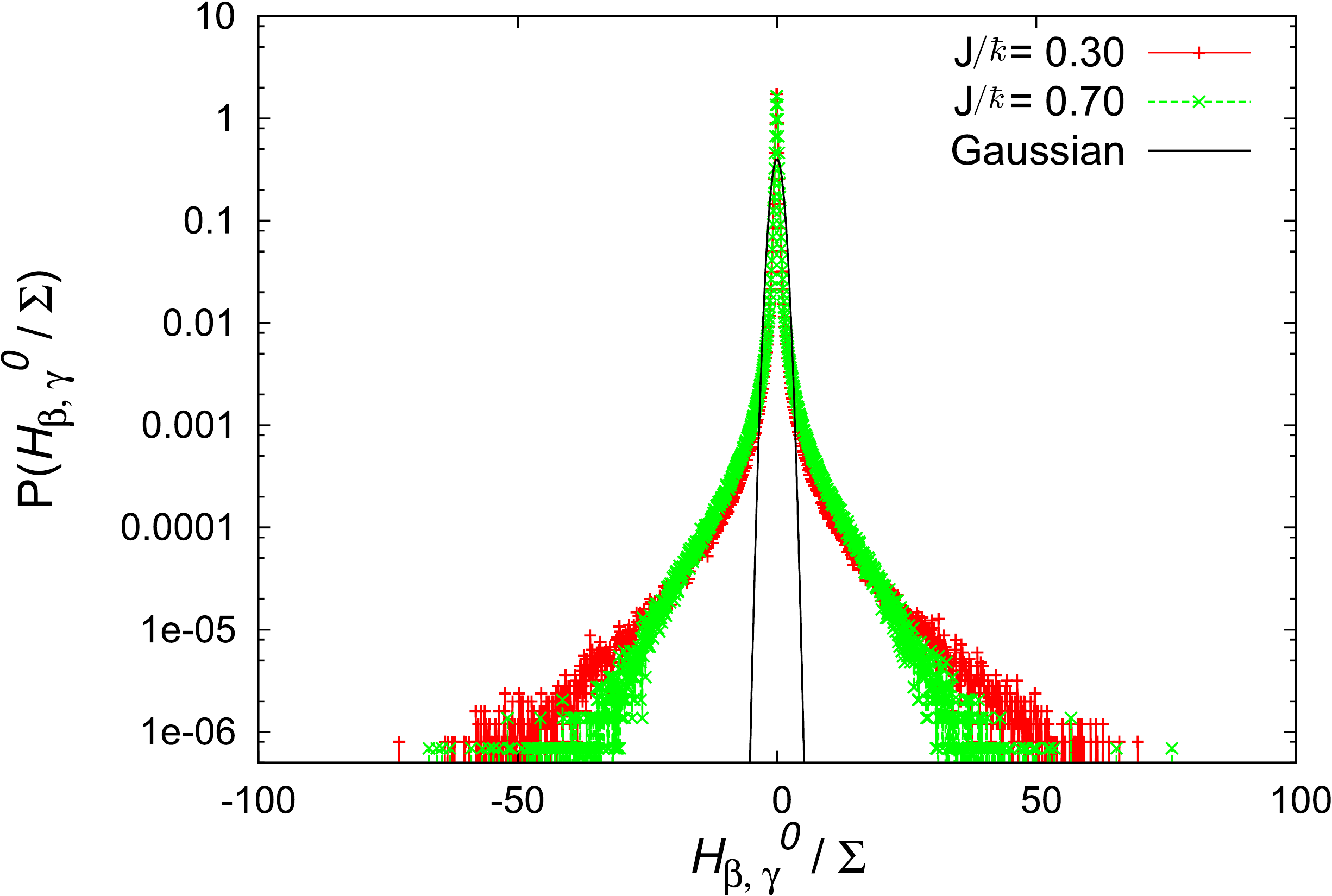}
\end{tabular}
\includegraphics[width=8cm]{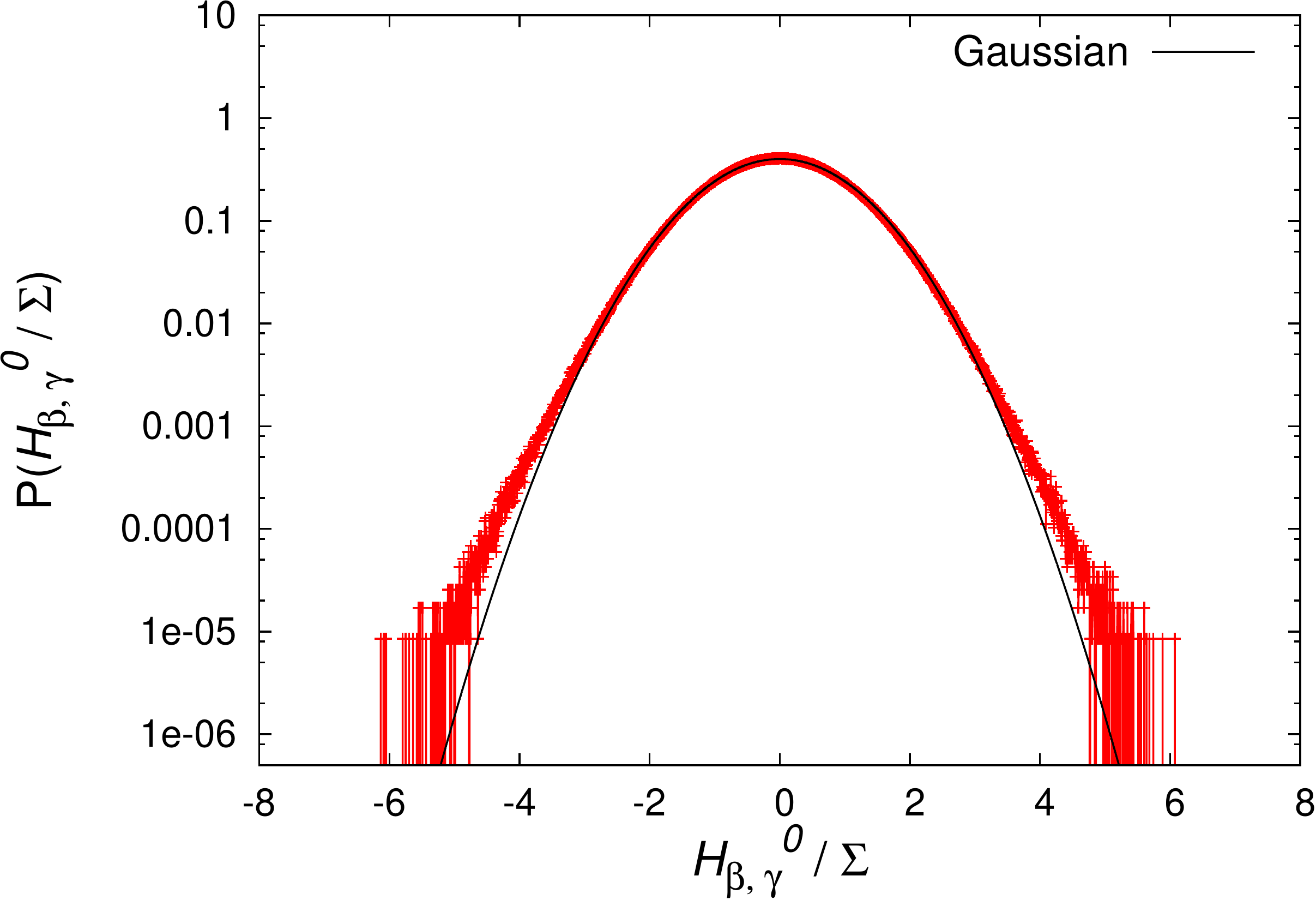}
\end{center}
\caption{Distributions normalized by the variance $\Sigma$ of the off-diagonal elements of $\mathcal{H}^{\,0}_{\beta\gamma}$. Numerical parameters: (left panel) $\jj=0.7$, (right panel) $\kk = 2.2$, (lower panel) $\kk=10.0$, $\jj=10.0$. $\kbar=5.0$, $L=3$, $M_{\rm max}=41$.}
\label{fig:distrib}
\end{figure}
\subsection{Level-spacing ratio analysis and {dynamical regimes}} \label{levelspacing:sec}
%
Here we consider the properties of the Floquet quasienergies. In particular, in order to understand if the dynamics is chaotic, we study the average level spacing ratio~\cite{oganesyan2007localization}. It is defined as
\begin{equation}\label{eq:r_ave}
\langle r \rangle = \frac{1}{D-2}\sum_{\beta=1}^{D-2}\frac{\mathrm{min}\{\lambda_\beta, \lambda_{\beta+1}\}}{\mathrm{max}\{\lambda_\beta, \lambda_{\beta+1}\}}\,,
\end{equation}
where $\lambda_\beta=\mu_{\beta+1}-\mu_\beta$.
If the dynamics is chaotic, the Hamiltonian should behave as a random matrix in the angular momentum basis:
The level spacings $\lambda_\beta$ obey the Circular Orthogonal Ensemble (COE) distribution and the average level spacing ratio is $\langle r \rangle\simeq0.5269$ (the Floquet operator belongs to the Circular Orthogonal Ensemble of symmetric unitary matrices~\cite{Rigol_PRX14,Haake:book,eynard}).  

On the opposite, a regular non-thermalizing behavior generically corresponds to a Poisson distribution~\cite{Berry375} of the $\lambda_\beta$; in this case the average level spacing ratio is $\langle r \rangle\simeq0.386$. These considerations are important for the energy absorption. As we extensively analyzed in~\cite{nostrolavoro}, a chaotic behavior corresponds to Floquet states delocalized in the angular momentum basis and to energy absorption. On the opposite, a regular behavior corresponds to localized Floquet states and then to dynamical localization.

In Fig.~\ref{fig:deloc_area}  we plot $\langle r \rangle$ as a function of $\kk$ for different values of $\jj$. In Fig.~\ref{fig:phase_diagram} we map the 
{regimes we observe in our model into the parameter space.}
We recognize 
the light-blue region on the right where $\mean{r}$ acquires the COE value and the energy increases up to a value scaling with the truncation. This is the chaotic dynamically delocalized region where the system heats up without a bound. This heating does not always correspond to full chaoticity and diffusive energy behavior: subdiffusion occurs in 
{the delocalized regime, close to its boundary}.
This is strictly reminiscent of the subdiffusion in space domain occurring in the delocalized phase near the transition to MBL~\cite{marko,Matthew}. Going on with our description of the 
diagram of Fig.~\ref{fig:phase_diagram}, we find on the left a red region which we define ``dynamically localized'' because we have numerically verified that the energy saturates after a transient to a value independent of the truncation [see an example in Fig.~\ref{fig:en_I_ratio}(b)]. This fact marks the presence of dynamical localization. Here $\mean{r}$ takes the Poisson value and the dynamics is regular-like. In between the localized and the delocalized region there is an intermediate regime where $\mean{r}$ has a value in between Poisson and COE. Here it is not easy to characterize the energy dynamics and we postpone the analysis of this regime to a future publication. 
%
\begin{figure}
  \begin{center}
%
%
     \includegraphics[width=8cm]{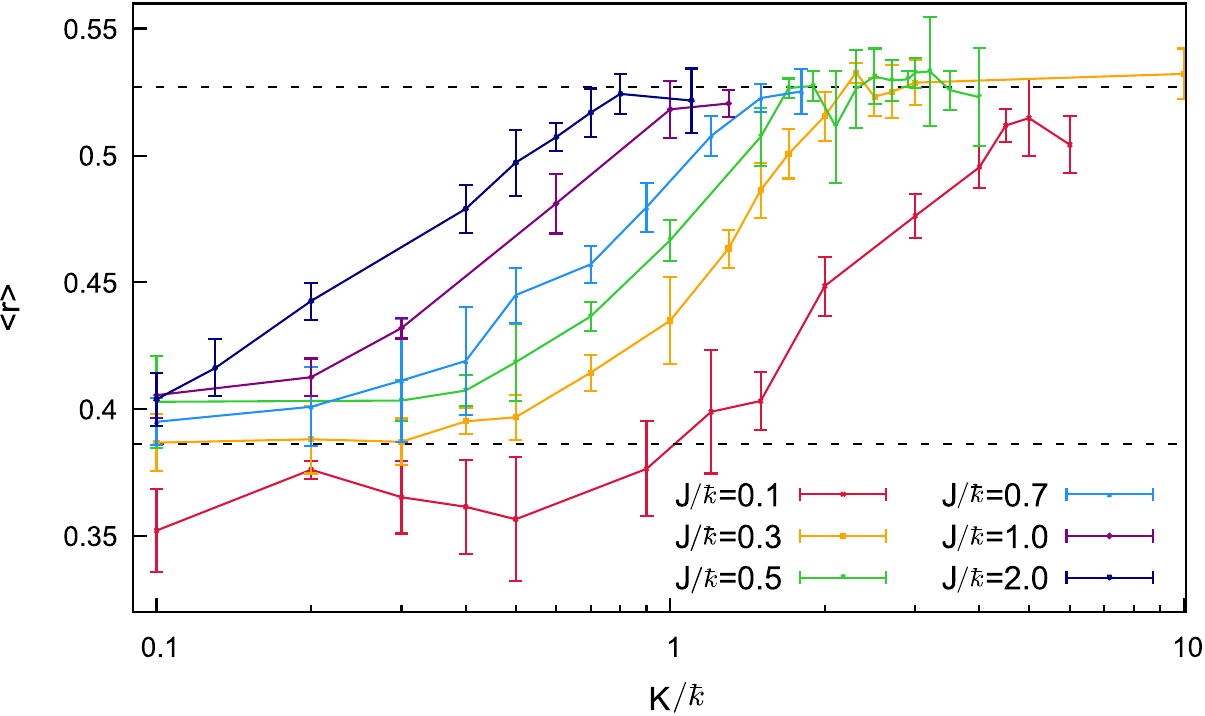}
%
%
  \end{center}
  \caption{\label{fig:deloc_area} The behavior of the average ratio $\langle r \rangle$ against $\kk$ for different values of $\jj$. The errorbars are obtained by averaging over the values obtained with different values of $M$. In this plot the maximum value of $M$ which has been used is $M_{\rm max}=39$. Other parameters: $L=3$, $\kbar=5.0$.}
\end{figure}
\subsection{Inverse participation ratio
 } \label{IPR:sec}
In order to explore the eigenstate thermalization breaking related to subdiffusion we have to consider the localization properties of the Floquet states. This can be realized by looking at the average Inverse Participation Ratio (IPR)~\cite{Wegner1980,Edwards_JPC72} in the angular momentum basis, defined as
\begin{equation}\label{eq:IPR_scaling}
\mean{\rm IPR}=\frac{1}{D}\sum_\beta \sum_{\bf m} |\langle\psi_\beta|{\bf m}\rangle|^{2q}\,,
\end{equation}
where $q=2$ is an exponent whose meaning will be clarified later. Given a state $|\psi\rangle$ which is uniformly delocalized  over all the states of the angular momentum basis, one finds that its IPR satisfies $\mathrm{IPR}_\psi\sim 1/D$. 

Let us focus on the delocalized region of the parameter space [see Fig.~\ref{fig:phase_diagram}]. We remind that we are in the truncated Hilbert space whose dimension is $D=M^L/(4L)$. If we suppose that there is eigenstate thermalization, then all the Floquet states are locally equivalent to the $T=\infty$ density matrix. They behave as random states and are fully delocalized, we can therefore infer that they should satisfy $\mean{\mathrm{IPR}}\sim 1/D$ as well. We find, instead, $\mean{\mathrm{IPR}}\sim 1/D^\delta$ with $\delta<1$. This fact occurs in a wide range of parameters inside the delocalized region, as we can see in Fig.~\ref{fig:ipr_vs_K_J_03_deloc}. In Fig.~\ref{fig:ipr_vs_K_J_03_deloc} (a) we show some example of the scaling of the IPR with $M$ at a fixed value of $\jj$ and some values of $\kk$ chosen in the dynamically-delocalized {regime},
 while in panel (b) the corresponding exponent $\delta$ is plotted versus $\kk$ (dark line). From a physical point of view this fact marks the breaking of eigenstate thermalization near the boundary of the delocalized region which we have already discussed from a different point of view in Sec.~\ref{risultati:sec}. 

Breaking of eigenstate thermalization implies that the Floquet states are not all equivalent, so we expect large fluctuations from one state to the other. In order to estimate the fluctuations in the IPR, we consider the scaling of the logarithmic average of the IPR, defined as 
\begin{equation}\label{eq:log_average_def}
\mathrm{IPR\,}_{\rm log} = \mathrm{exp}\langle \mathrm{ln}\, \mathrm{IPR}\rangle\,,
\end{equation}
where $\mean{(\ldots)}=\frac{1}{D}\sum_\beta(\ldots)_{\beta}$. We find a scaling of the form $\mathrm{IPR\,}_{\rm log}\sim 1/D^\eta$; we plot the scaling exponent $\eta$ versus $\kk$ in Fig.~\ref{fig:ipr_vs_K_J_03_deloc} (b) (lighter curve). We see that the exponents $\delta$ and $\eta$ are similar and near 1 for high values of $\kk$. Here all the Floquet states are similar (small fluctuations) and obey eigenstate thermalization (full delocalization). For smaller $\kk$, on the opposite, the exponents are different and smaller than 1. Here eigenstate thermalization is broken {and large fluctuations in the IPRs of the Floquet states emerge}. {We would like to emphasize that the  eigenstate thermalization breaking witnessed by $\delta\,,\eta<1$ in Fig.~\ref{fig:ipr_vs_K_J_03_deloc} and the subdiffusion with $\alpha<1$ occur in the same parameter range (see Fig.~\ref{fig:pl_exponents}): Further research will be devoted to investigate the relation between these two results.}

In order to better understand the origin of the fluctuations of the Floquet IPRs, we have focused on their distributions. To this purpose we have considered a fixed value of $\jj$ and computed the distributions for some values of $\kk$, as shown in Fig.~\ref{fig:IPR_distr}(a): The distributions exhibit a power-law tail for smaller values of $\kk$ which tends to disappear as $\kk$ is increased. This observation is in agreement with the behavior of the exponents $\delta$ and $\eta$ discussed above: a long tail means large fluctuations and then $\delta\neq\eta$. Moreover, a long tail means coexistence of more localized and more delocalized states with states not completely localized. A last remark regards the behavior of the distributions with $M$, which is shown in Fig.~\ref{fig:IPR_distr}(b). In the presence of fluctuations [$\delta\neq\eta$ -- left panel of Fig.~\ref{fig:IPR_distr}(b)] a small fraction of localized states persists as $M$ is increased, as it emerges from the right extreme of the power law-tail. Differently, in absence of fluctuations and with $\delta=1$, all the states are delocalized and the whole distribution shifts to lower values of the ${\rm IPR}$ as $M$ is increased [Fig.~\ref{fig:IPR_distr}(b)].

 {It is worth noting that similar scaling properties of the IPRs have been found
 at the Anderson transition point of  a three-dimensional disordered lattice \cite{PhysRevLett.79.1913,0305-4470-19-8-004,PhysRevB.84.134209} and across the MBL transition \cite{PhysRevB.96.104201}: 
The eigenstates exhibit  an anomalous scaling of the probability distribution momenta \cite{chalker1996spectral} and they are said to have a multifractal structure. In our analysis,
 we focused on the averaged second momentum,
 namely the $\mean{\rm IPR}$. By computing the exponent $\delta$ for higher momenta ($q>2$),
 a set of fractal dimensions  can be obtained for characterizing a single state. Multifractal states  are also characterized through their probability distribution correlation functions and their level spacing distributions. The mapping existing between this rotors model and a disordered $L$-dimensional lattice~\cite{nostrolavoro} suggests the existence of multifractal properties also in our case. This multifractal analysis, which may help understanding the intermediate region in Fig.~\ref{fig:phase_diagram}, is left for future work.
 }
%
\begin{figure}
  \begin{center}
    \begin{tabular}{c}
%
   \vspace{10pt}
     \includegraphics[width=8cm]{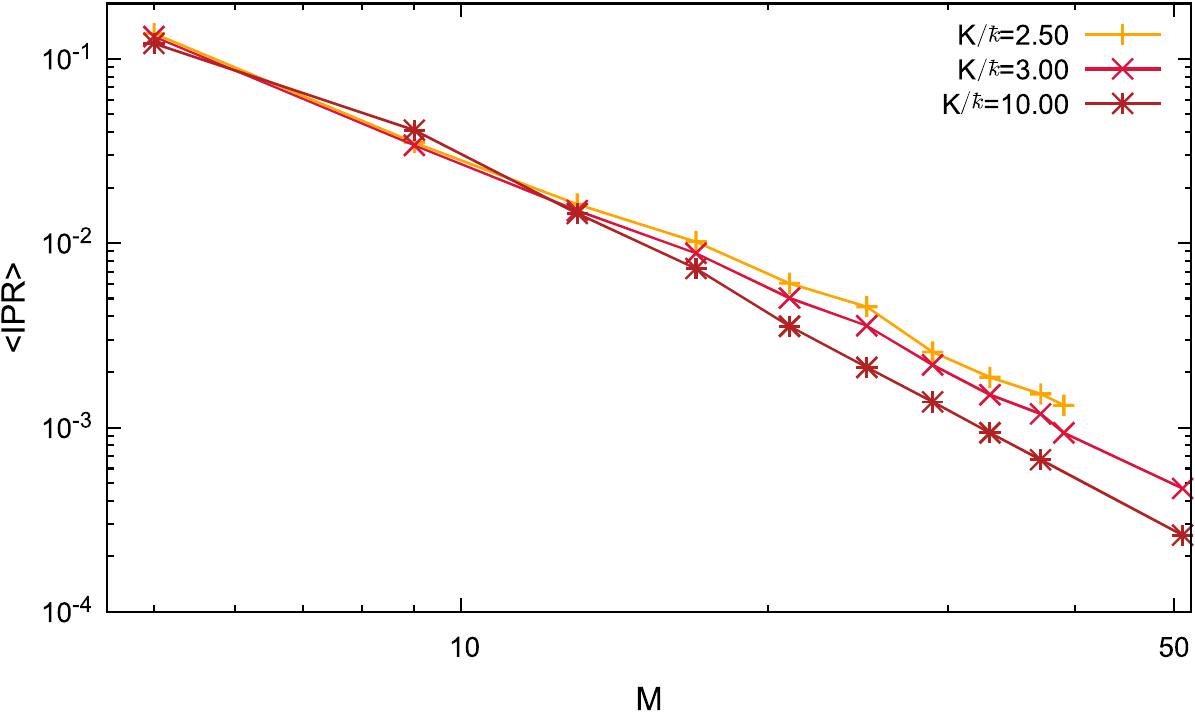}\\
     \pgfputat{\pgfxy(-4.3,5)}{\pgfbox[left,top]{\footnotesize (a)}}\\
     \includegraphics[width=8cm]{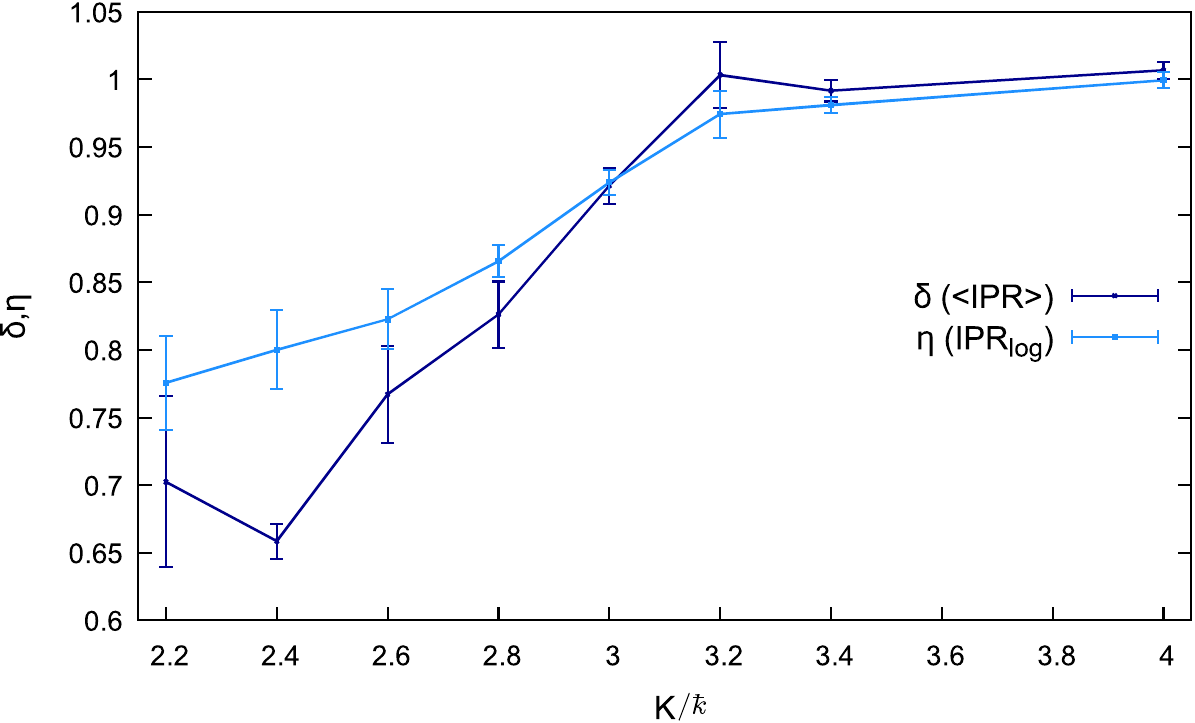}\\
     \pgfputat{\pgfxy(-4.3,5)}{\pgfbox[left,top]{\footnotesize (b)}}\\
%
    \end{tabular}
  \end{center}
  \caption{\label{fig:ipr_vs_K_J_03_deloc} (a) The behavior of the average IPR vs $M$ is plotted for $\jj=0.3$ and $\kk=2.5,\,3.0,\,10.0$ (from lighter to darker colors). For this values the system is delocalized and the $\mean{\rm IPR}$ scales as a power law with the dimension of the Hilbert space. (b) Exponent $\delta$ relative to the scaling of the $\mean{\rm IPR}$ (darker curve) and $\eta$ relative to the scaling of the logarithmic average (lighter curve). Other parameters: $L=3$, $\kbar=5.0$.}
\end{figure}

\begin{figure}
     \begin{center}
      \begin{tabular}{c}
     \vspace{10pt}
       \includegraphics[width=8cm]{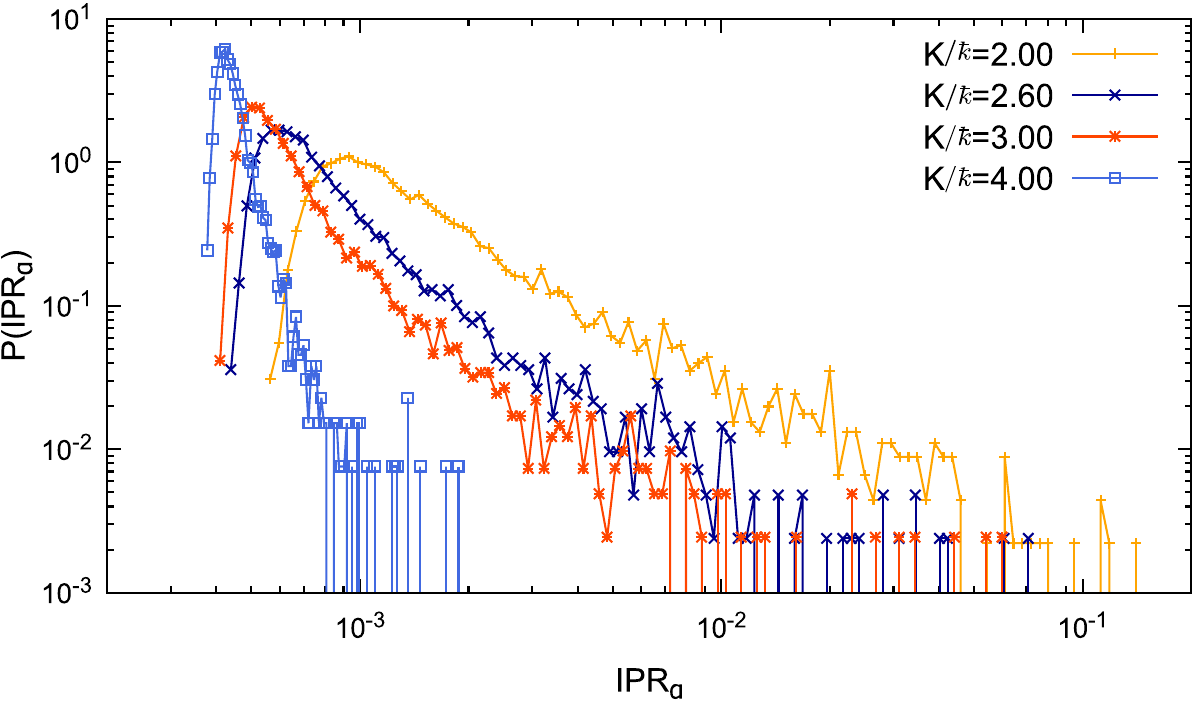}\\
       \pgfputat{\pgfxy(-5.5,5)}{\pgfbox[left,top]{\footnotesize (a)}}\\
       \includegraphics[width=10cm]{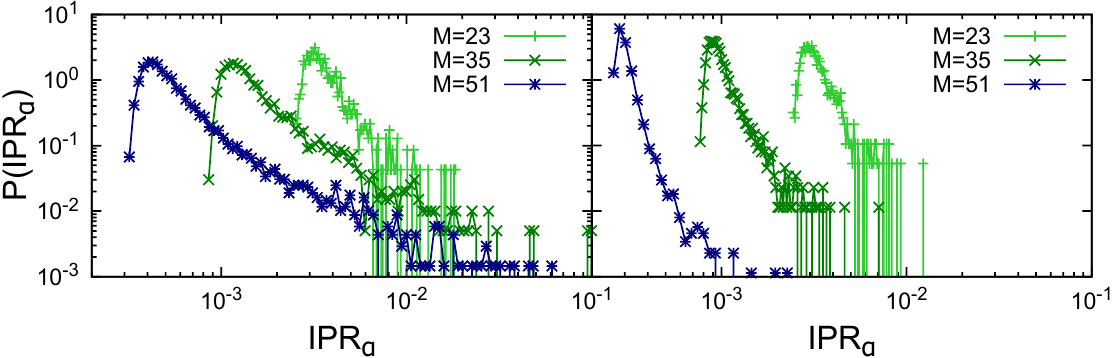}\\
        \pgfputat{\pgfxy(-5.5,4)}{\pgfbox[left,top]{\footnotesize (b)}}
       \end{tabular}
     \end{center}
  \caption{\label{fig:IPR_distr}(a) Distributions of the IPR at a fixed value of $\jj$ and different values of $\kk$. The power law tails disappear as $\kk$ increases and the system becomes fully ergodic. (b) Distributions of the ${\rm IPR}_\alpha$ at different values of $M$ in a case in which there are fluctuations (left, $\kk=2.6$) and in a case in which there are not (right, $\kk=4.0$). Parameters values: $\jj=0.3$, $L=3$, $\kbar=5.0$.}
\end{figure}
\subsection{Asymptotic behavior of the energy} \label{asymptotic:sec}
The asymptotic behavior of the kinetic energy is governed by the diagonal matrix elements $\mathcal{H}^{\,0}_{\beta\beta}$: At long times, the system reaches the infinite-time  averaged kinetic energy, defined as
\begin{equation}\label{eq:diag_H0}
E_{n\to\infty}=\lim_{\mathcal{T}\to\infty}\frac{1}{\mathcal{T}}\sum_{n=0}^\mathcal{T}E(n)=\sum_\beta |\psi_\beta|^2 \mathcal{H}^{\,0}_{\beta\beta}\,.
\end{equation}
The value of $E_{n\to\infty}$ is independent of $M$ in the case of dynamical localization: The initial state evolves until it reaches an asymptotic state which is localized in the angular momentum space. On the other side, when the dynamics is ergodic the system is expected to reach the so called infinite-temperature state, defined as 
\begin{equation}\label{eq:inf_state}
\rho_{T\rightarrow\infty}=\frac{1}{D}\sum_\beta |\psi_\beta\rangle\langle\psi_\beta|\,.
\end{equation}
This state is the equivalent, in the infinite-temperature case, of the equilibrium thermal state which is reached by time-independent ergodic systems at a given temperature $T$. The corresponding expectation value of the energy, obtained by taking $E(T=\infty)=\mathrm{Tr}[\hat H_0\, \rho_{T\rightarrow\infty}]/L$, is:
\begin{equation} \label{eq:inf_temp_state}
E(T=\infty) = \frac{\kbar^2}{2\,LD} \sum_{\bf m} |{\bf m}|^2\,.
\end{equation}
In the  case of delocalized dynamics the ratio $I(M)$ defined as
\begin{equation}
I(M)=\frac{E_{n\rightarrow\infty}}{E(T=\infty) }
\end{equation}
should be $O(1)$ with respect to the truncation $M$. {Differently, in the dynamically-localized {regime}
 the averaged, infinite-time energy is constant with respect to $M$, while $E(T=\infty)\sim M^2$, so we have $I(M)\sim M^{-2}$~\footnote{This relation comes by observing that the infinite temperature energy of a single rotor can be re-written as $\frac{\kbar^2}{2\,M} \sum_{m=-m_{max}}^{m_{max}} m^2$.}.
}
\begin{figure}
     \begin{center}
     \vspace{10pt}
     \begin{tabular}{c}
       \includegraphics[width=8cm]{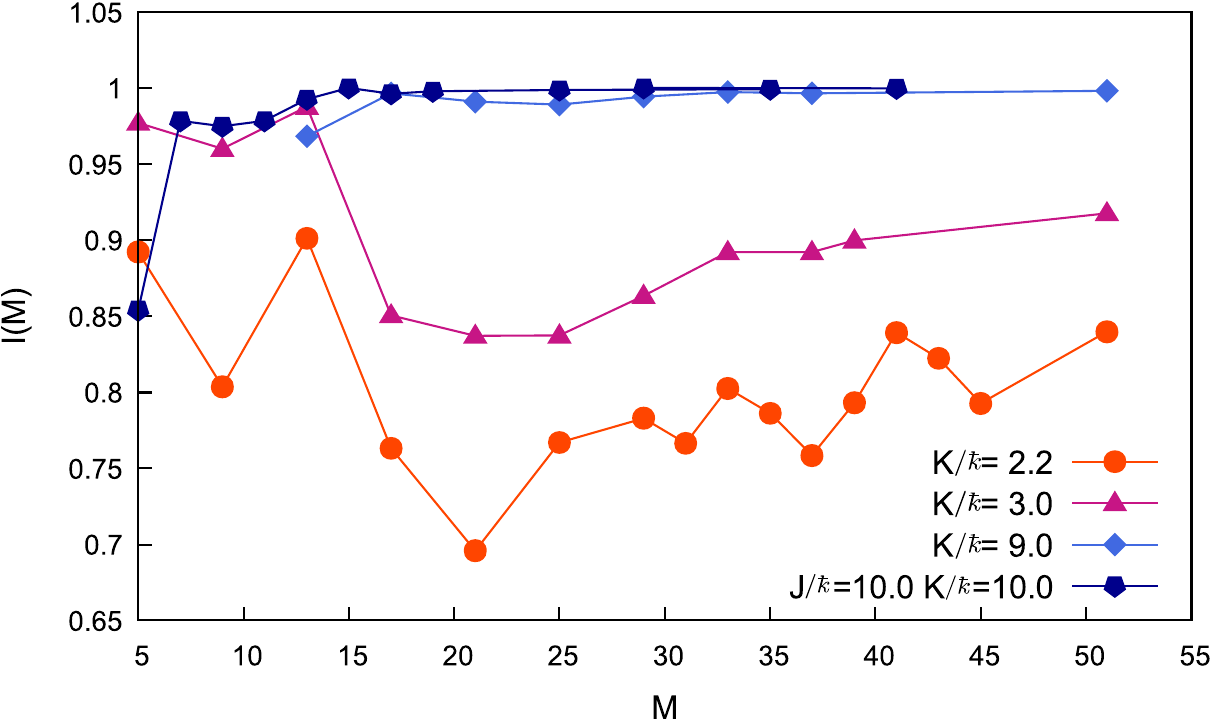}\\
       \pgfputat{\pgfxy(-5.,5)}{\pgfbox[left,top]{\footnotesize (a)}}\\
        \includegraphics[width=8cm]{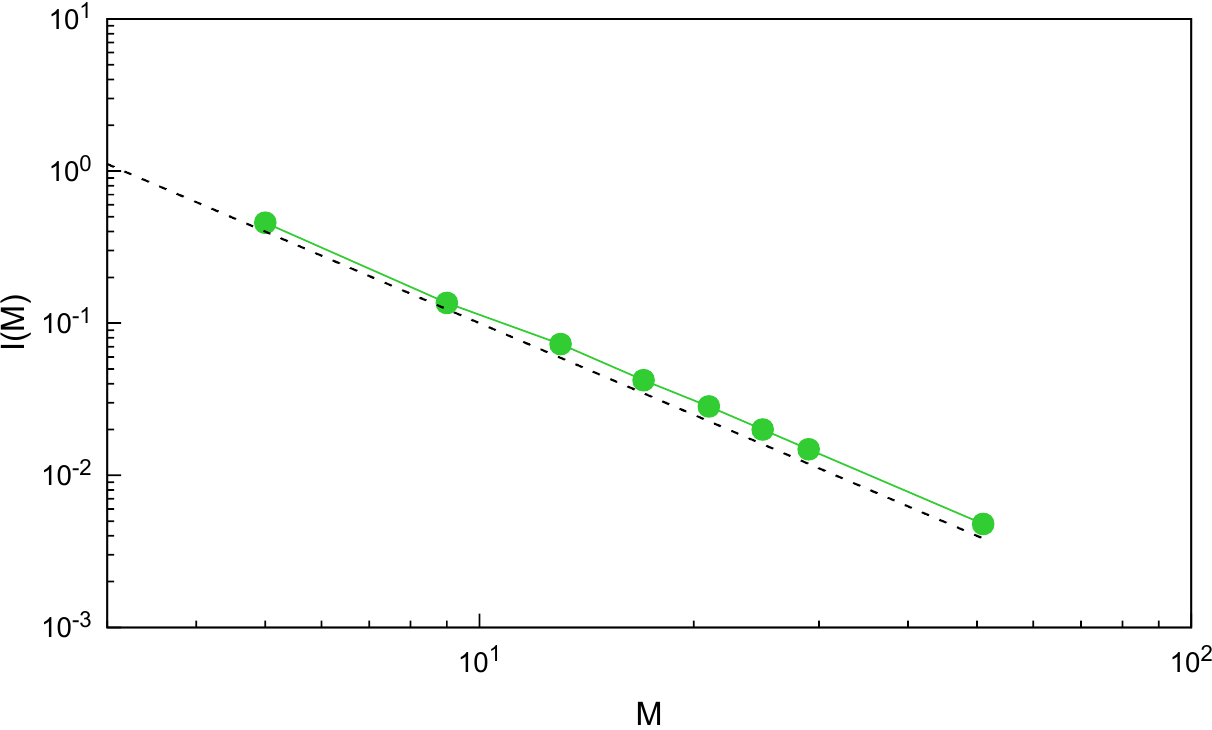}\\
         \pgfputat{\pgfxy(-5.,4)}{\pgfbox[left,top]{\footnotesize (b)}}
        \end{tabular}
     \end{center}
  \caption{\label{fig:en_I_ratio} (a) The ratio $I(M)$ is plotted for several values of $\kk$ at $\jj=0.3$ and in the case $\kk=10.0$ and $\jj=10.0$. Note that the ratio is $O(1)$ with respect to $M$. (b) I(M) plotted for $\kk=0.4$ and $\jj=0.3$ in the dynamically-localized {regime}.
 The points follow a power law with slope $\alpha=-1.94 \pm 0.016$ while the dashed curve has the expected slope -2. Other parameters: $L=3$, $\kbar=5.0$. 
  }
\end{figure}
{
 In Fig.~\ref{fig:en_I_ratio} we plot the ratio $I(M)$ at a fixed value of $\jj$ for some choices of $\kk$ in the dynamically-delocalized {regime}
 (a) and in the localized one (b). In the latter case we find the expected power-law behavior of $I(M)$; in the former case we find that $I(M)$ does not change significantly with $M$ in the interval we can access with simulations.} Notice that in case of ergodicity one would expect, in particular, $I(M)=1$, and this occurs for the thermalizing case $\kk=\jj=10$. In the other subdiffusive cases, we observe that $I(M)$ increases towards $1$ as $\kk$ is increased in the range of accessible values of $\kk$. Nevertheless, due to  numerical limitations, we cannot distinguish whether $I(M)$ tends to 1 for larger values of $M$, independently on $\kk$.
{
The diffusion and thermalization occurring for large kick parameters may suggest that in the fully ergodic regime each rotor evolves 
as if it interacts with an external bath. Hence some of the phenomenology we observe may have connections with the results in \cite{PhysRevLett.83.61, d2001quantum, ammann1998quantum}
}

We remark the strong dependence of $I(M)$ on $\kk$, in opposition with the behavior of the subdiffusion exponent $\alpha$ (see Fig.~\ref{fig:pl_exponents}). Nevertheless, we see that $I(M)$ is significantly smaller than 1 in the same interval where there is subdiffusion and no eigenstate thermalization [see Figs.~\ref{fig:pl_exponents} and~\ref{fig:IPR_distr}(b)], marking the connection between these phenomena.

\section{Analysis of the random-matrix model} \label{randommat:sec}
We have seen in the previous section that the operator $\hat{H}_0$ does not look like a random matrix in the Floquet basis, as the anomalous distributions of $\mathcal{H}^{\,0}_{\beta\gamma}$ of Fig.~\ref{fig:distrib} testify. Here we build a model which can reproduce these distributions. The rationale is the following. In the chaotic case, the operators look like random matrices in the Floquet state basis~\cite{Peres_PRA84}. On the opposite, in the localized case the Hamiltonian ($\hat{H}_0$) looks like a banded random matrix~\cite{Molinari:banded,mario_jpa97} and one can argue that the operators showing localization appear as banded random matrices in the Floquet basis (otherwise they would be delocalized). We construct a model interpolating between these two conditions. It is a random matrix where the elements are Gaussian distributed, but the variance of this distribution depends on the position inside the matrix and gets smaller as the distance from the diagonal is increased. In particular, it depends as a power law on the distance from the diagonal, so we assume that $\mathcal{H}^0_{\beta\gamma}$ for $\gamma\neq\beta$ is a Gaussian random variable with variance
\begin{equation}\label{decay:eqn}
  \sigma_{\beta,\gamma}=\sigma_{|\beta-\gamma|}=\frac{1}{|\beta-\gamma|^b}\,,
\end{equation}
with $b$ some real non-negative number. In the limit $b=0$ we recover the standard random-matrix behavior, while in the limit $b\to\infty$ we move towards a banded random matrix behavior.

In order to show the validity of our model, we can use it to fit the distributions of $\mathcal{H}^0_{\beta\gamma}/\Sigma$ (Fig.~\ref{fig:distrib}) obtained through exact diagonalization. To do the fit, we adjust the parameter $b$ in order to numerically minimize the quantity
\begin{equation}
  d_{\rm log}(b)=\int_{-\infty}^\infty\ud x\left|\log(P(x))-\log(P_{\rm model}^{(b)}(x))\right|\,.
\end{equation}
We perform some plots of $d_{\rm log}(b)$ versus $b$. For small values of $b$, our numerics gives us distributions restricted to a too narrow interval, that's why we have a non-physical increase of $d_{\rm log}(b)$ (we do not show this interval of $b$ in Fig.~\ref{fig:distance}). For larger values of $b$ we find a physical minimum. 
The minimum we find is very shallow and we cannot clearly determine it, being overwhelmed by fluctuations (Fig.~\ref{fig:distance}). {Nevertheless, we find values of $b$ for which  the agreement between the distributions resulting from ED and those from this model is good (see Fig.~\ref{agreement3:fig})}. Notice the very clear fluctuations at large deviations, giving rise to the wigglings of $d_{\rm log}$. From our results, we see that $b$ is closer to 0 for large $K$. This is in agreement with the physical expectation that these cases are more chaotic and then closer to a pure random matrix condition.

%
\begin{figure}
\vspace{10pt}
\begin{center}
\begin{tabular}{c}
  \includegraphics[width=8cm]{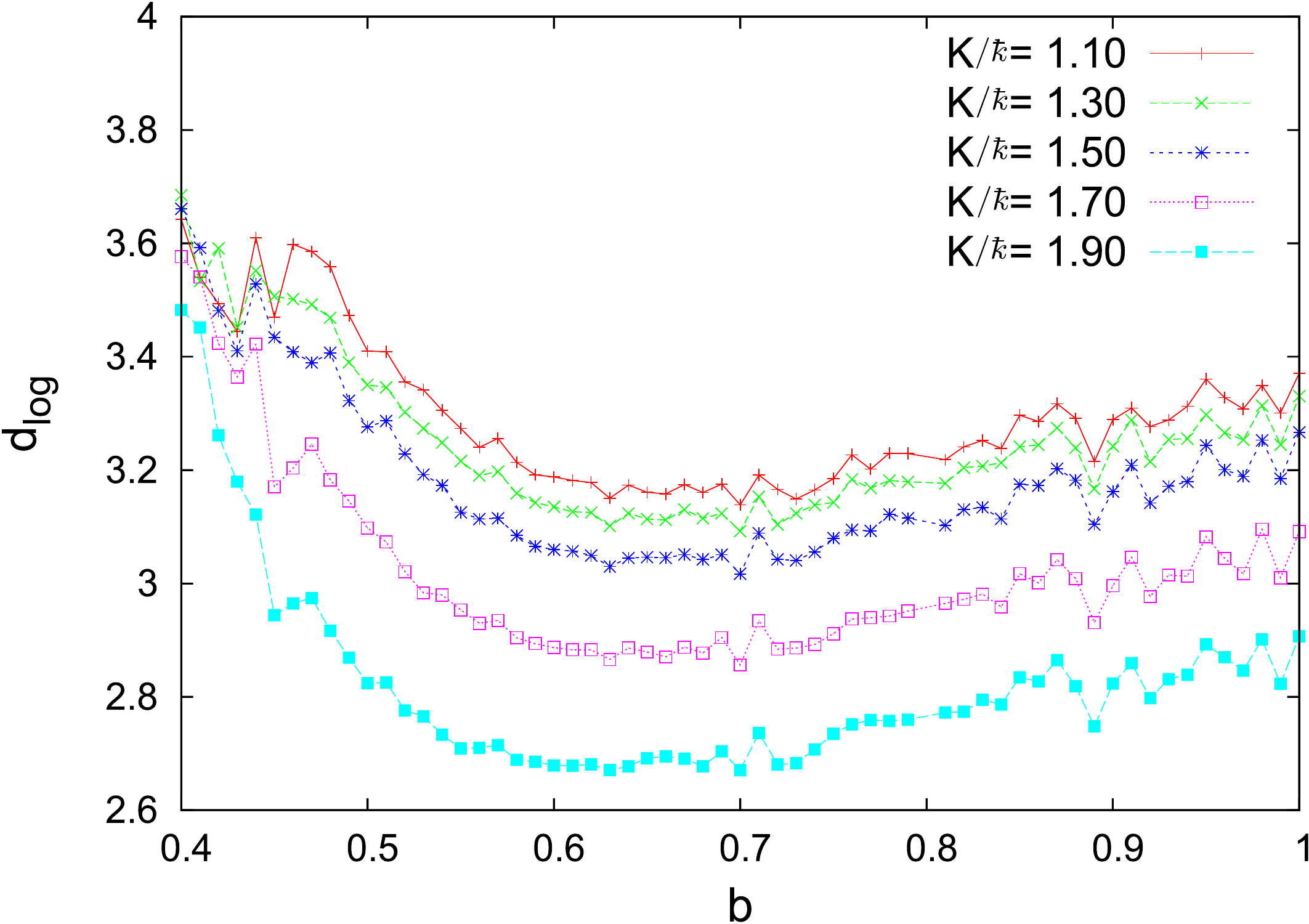}
\end{tabular}
\end{center}
\caption{$d_{\rm log}$ versus $b$. Numerical parameters: $\jj=0.7$, $\kbar=5.0$, $L=3$, $M_{\rm max}=41$.}
\label{fig:distance}
\end{figure}
%
%
%
%
%
%
%
\begin{figure}
  \begin{center}
   \begin{tabular}{cc}
%
      \hspace{-1cm}\includegraphics[width=8cm]{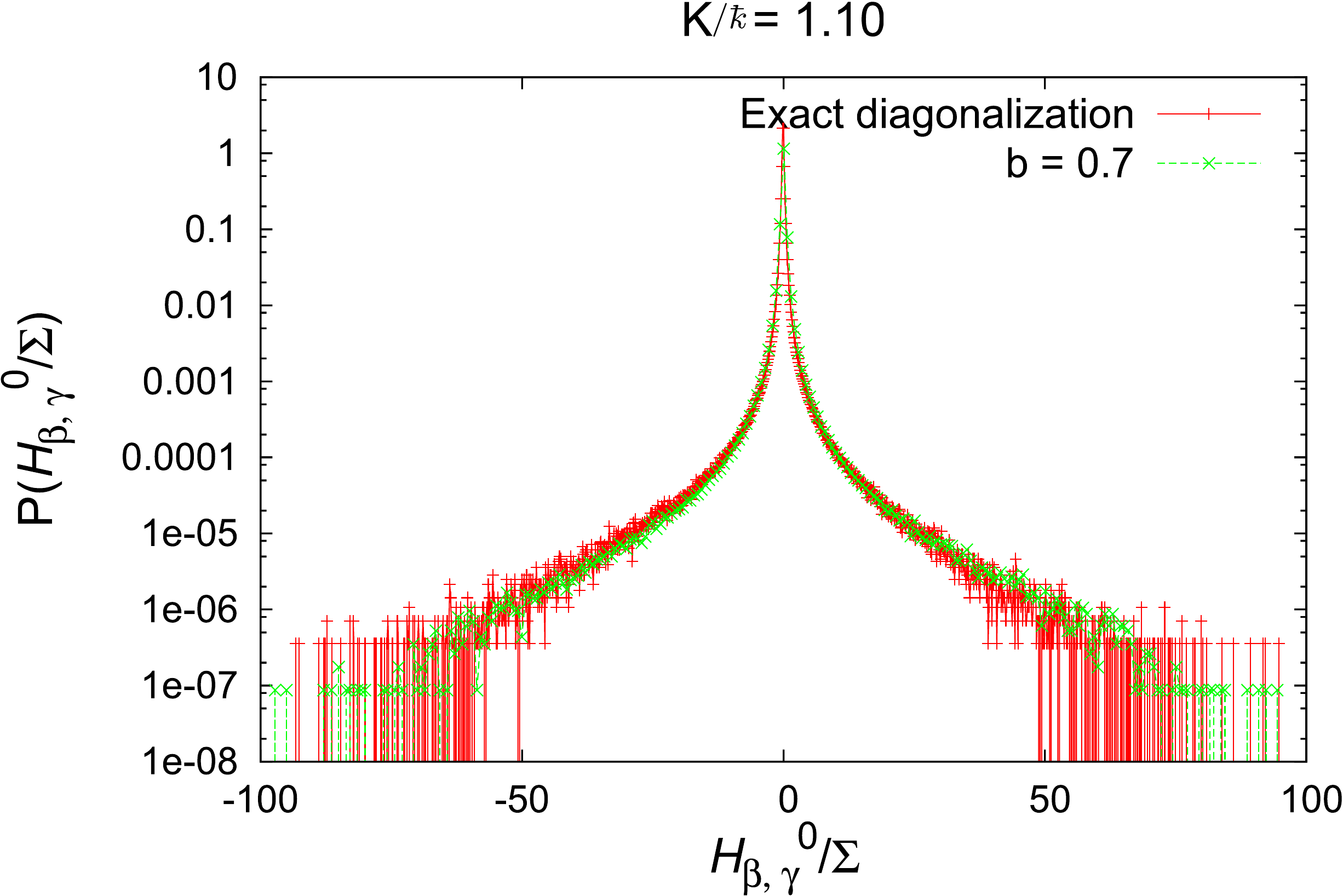}&
      \hspace{0cm}\includegraphics[width=8cm]{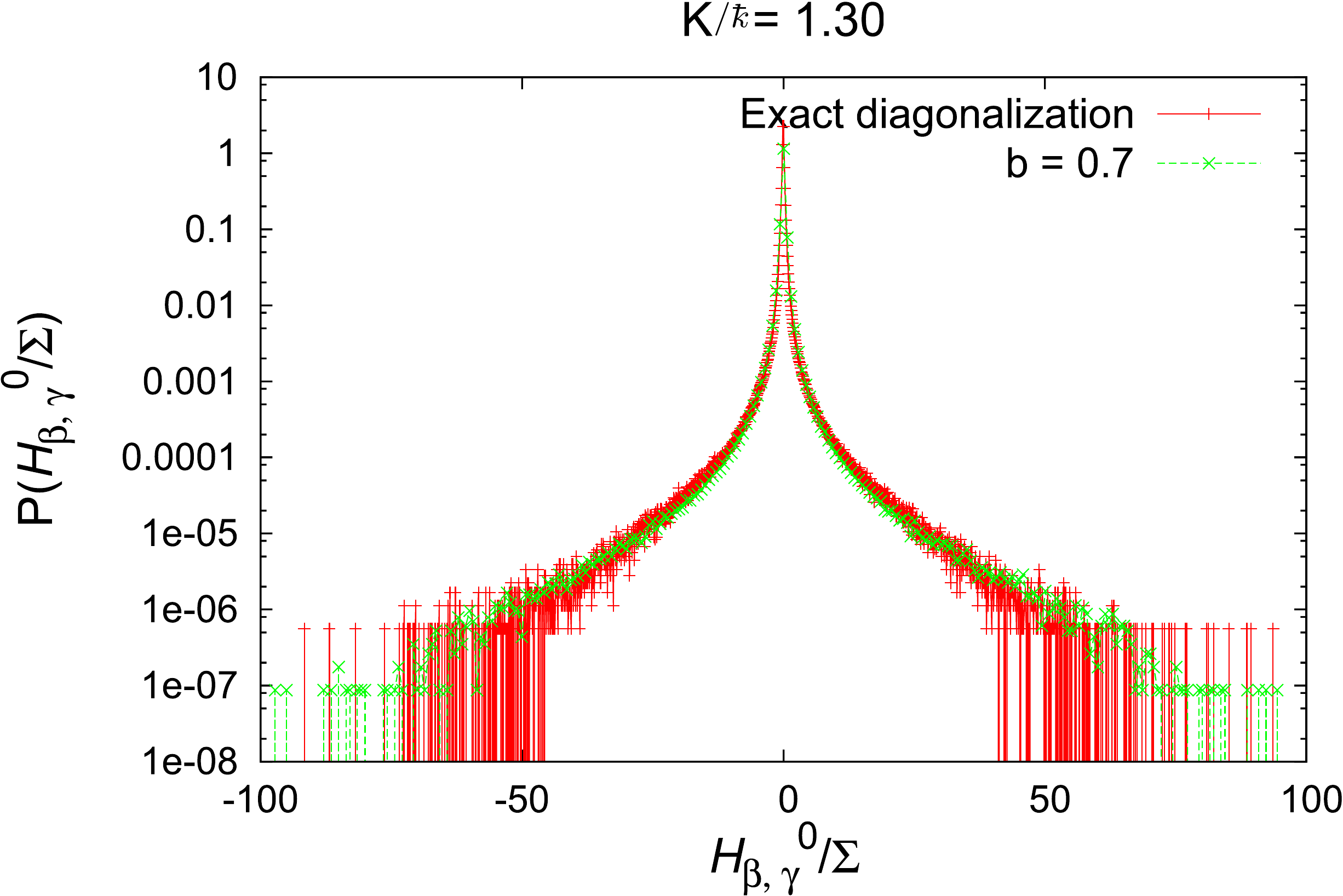}\\ \\
      \hspace{-1cm}\includegraphics[width=8cm]{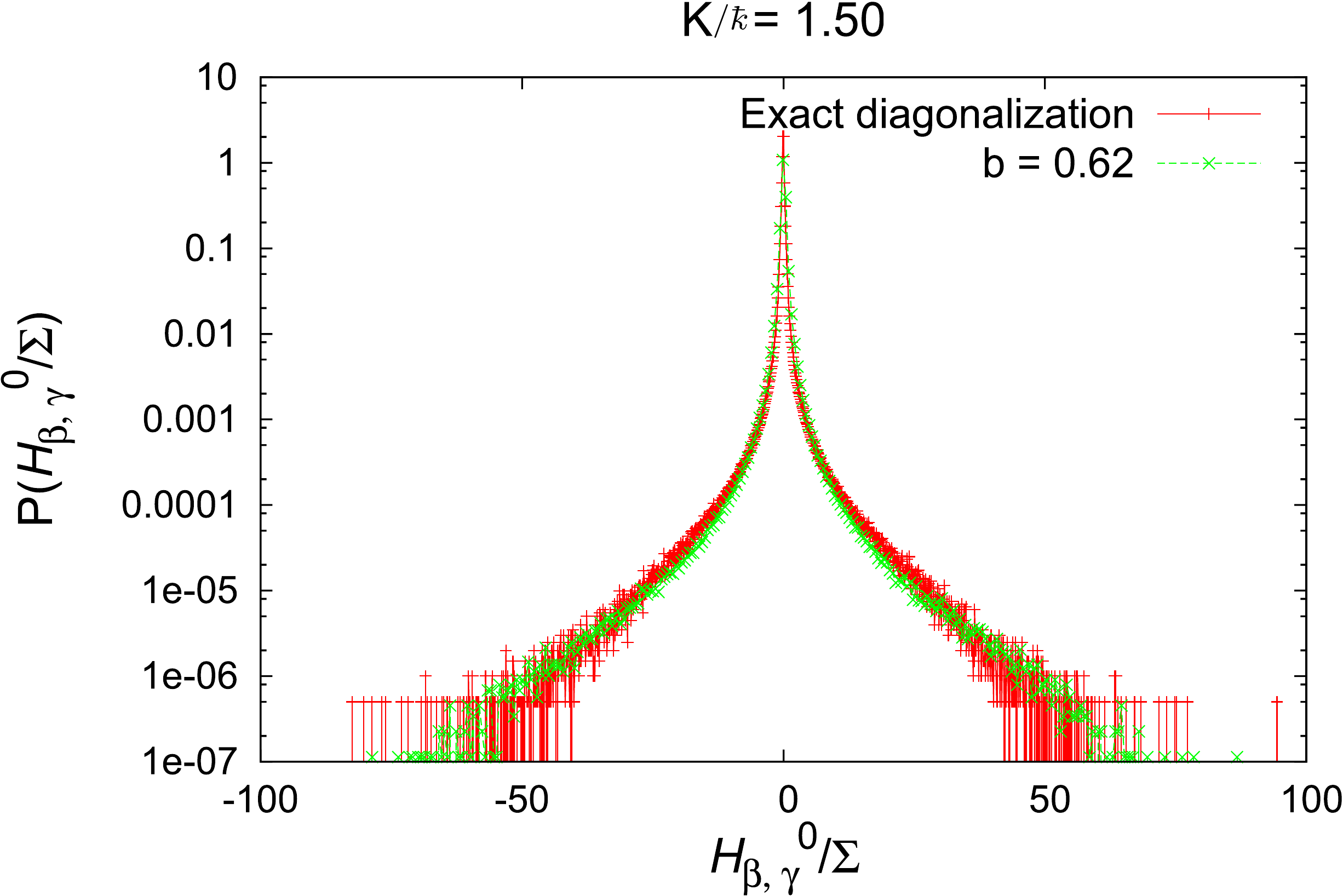}&
      \hspace{0cm}\includegraphics[width=8cm]{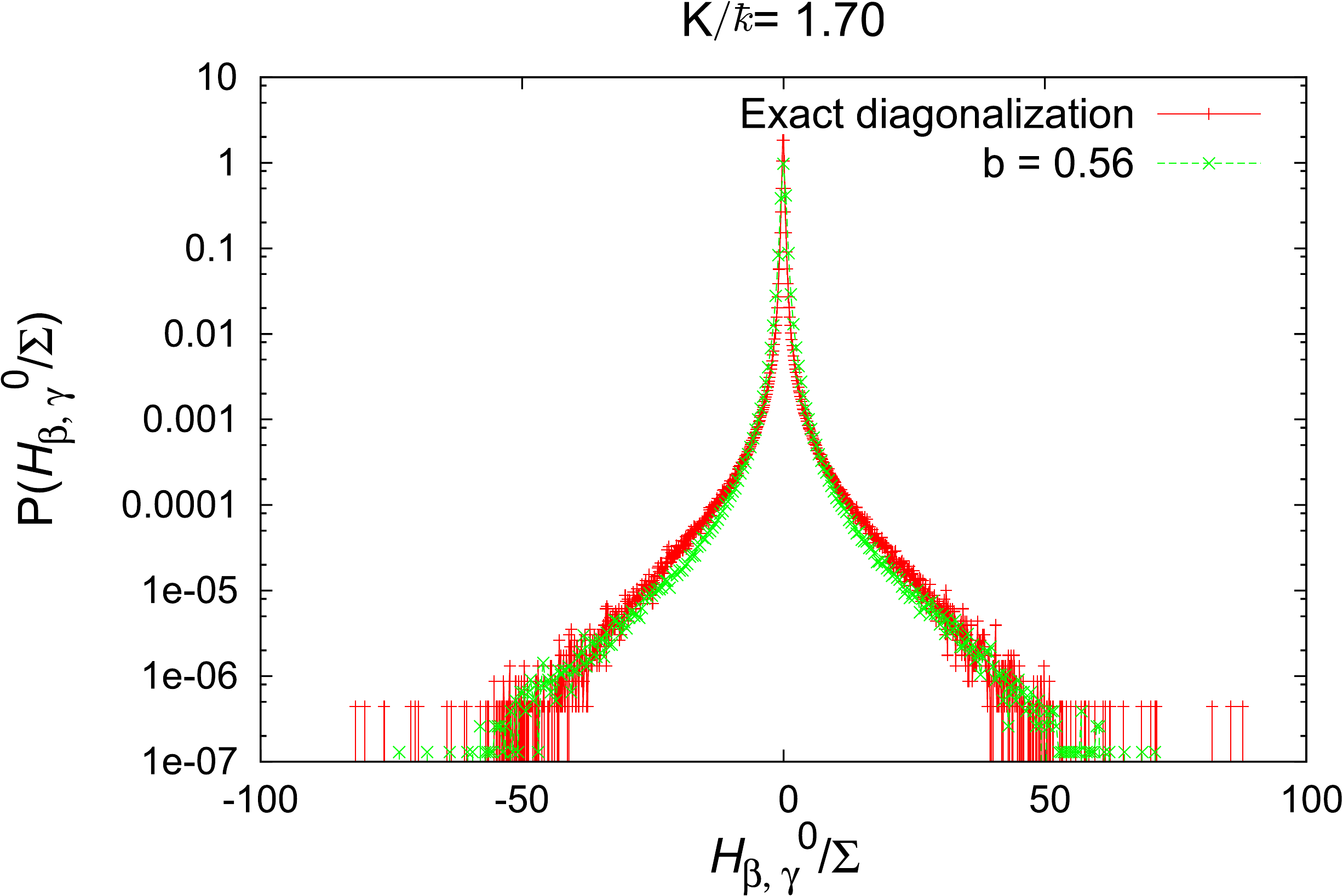}
   \end{tabular}
\vskip 12pt
   \hspace{0cm}\includegraphics[width=8cm]{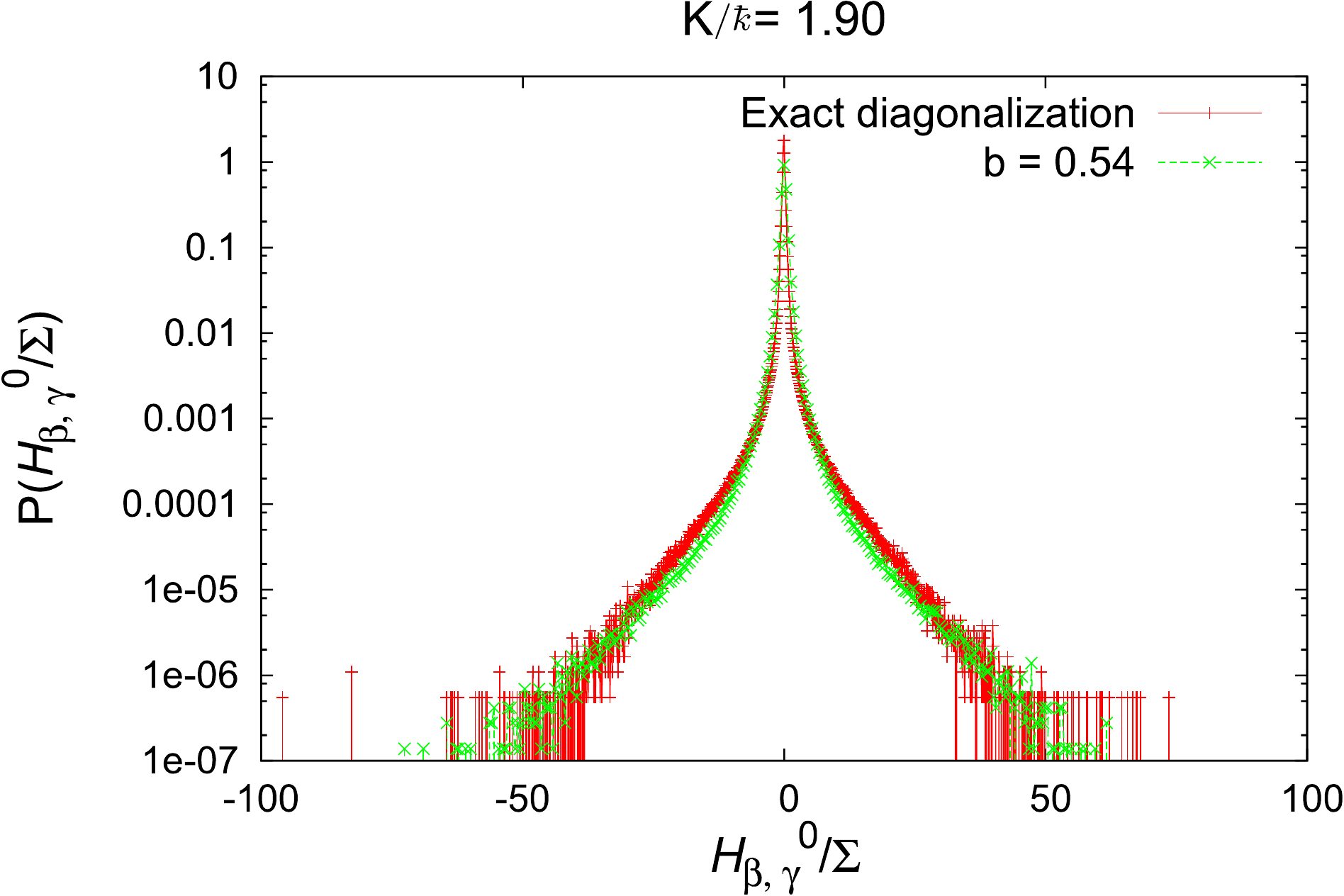}
  \end{center}
 \caption{Comparison of the exact-diagonalization distributions and the ones of the effective random-matrix model of Eq.~\eqref{decay:eqn} with appropriate choice of $b$. Notice the quite good agreement. Numerical parameters: $\jj=0.7$, $\kbar=5.0$, $L=3$, $M_{\rm max}=41$.}
 \label{agreement3:fig}
\end{figure}
%
%
\section{Conclusions} \label{conc:sec}
In conclusion we have studied the energy subdiffusion in an interacting quantum kicked rotors model. 
We have noticed that this is a purely quantum phenomenon and, through a numerical analysis, we have mapped 
{the different dynamical regimes}
in the parameter space.
 We have considered the subdiffusion and the asymptotic properties of the energy. 
 About the latter, the absence of full thermalization in the truncated Hilbert space is strictly related to an anomalous 
 behavior of the Floquet states, marking the breaking of eigenstate thermalization. 
 Subdiffusion is associated to anomalous non-Gaussian distributions of the off-diagonal matrix elements of the energy in the Floquet basis. 
 These distributions are well described by a model of anomalous random matrices. 
 
 Future directions of research include the application of the anomalous random matrix model to subdiffusion in ergodic systems near the MBL transition~\cite{marko,Matthew,Luitz_PRL16,Roy_PRB18,foini2019eigenstate}. 
 {It would be also worth investigating the nature of the intermediate region: by exploiting the mapping between the rotors model and the disordered Anderson one, a possible way would be  analyzing the possible multi-fractal structure of the Floquet states.}
%

\ack{We acknowledge useful discussions with V.~Kravtsov and F.~Iemini.
SN  acknowledges SISSA and ICTP for supporting this work during his PhD; SN acknowledges partial support from the H2020 Project - QUANTUM FLAGSHIP - PASQUANS (2019-2022).
}


%

\section*{Bibliography}
\vspace{0.5cm}

\begin{thebibliography}{10}

\bibitem{Berry_regirr78:proceeding}
M.~V. Berry.
\newblock {Regular and Irregular Motion}.
\newblock In S.~Jorna, editor, {\em {Topics in Nonlinear Mechanics}}, volume~46, pages 16--120. Am.Inst.Ph., 1978.

\bibitem{Lichtenberg}
A.J. Lichtenberg and Lieberman M.A.
\newblock {\em Regular and Chaotic Dynamics}.
\newblock Springer-Verlag, $2^{nd}$ edition, 1992.

\bibitem{salamon2004kolmogorov}
D.~Salamon.
\newblock The Kolmogorov-Arnold-Moser theorem.
\newblock {\em Math. Phys. Electron. J}, 10(3):1--37, 2004.

\bibitem{Arnold-Avez:book}
V.I. Arnold and A.~Avez.
\newblock {\em {Ergodic problems of classical dynamics}}.
\newblock W. A. Benjamin, Inc., 1968.

\bibitem{konishi}
T.~Konishi and K.~Kaneko.
\newblock Diffusion in Hamiltonian chaos and its size dependence.
\newblock {\em Journal of Physics A: Mathematical and General}, 23(15):L715,
  1990.

\bibitem{chiri_vov}
B.V Chirikov and V.V. Vecheslavov.
\newblock Theory of fast Arnold diffusion in many-frequency systems.
\newblock {\em Journal of Statistical Physics}, 71:243, 1993.

\bibitem{Ata_ema}
Emanuele Dalla~Torre and Atanu Rajak.
\newblock Characterizations of prethermal states in periodically driven
  many-body systems with unbounded chaotic diffusion.
\newblock {\em arXiv}, page 1905.00031, 2019.

\bibitem{Oven}
O.~Howell, P.~Weinberg, D.~Sels, A.~Polkovnikov, and M.~Bukov.
\newblock Asymptotic pre-thermalization in periodically driven classical spin
  chains.
\newblock {\em Phys. Rev. Lett.}, 122:010602, 2019.

\bibitem{Shirley_PR65}
J.~H. Shirley.
\newblock Solution of {S}chr{\"o}dinger equation with a Hamiltonian periodic in
  time.
\newblock {\em Phys. Rev.}, 138:B979, 1965.

\bibitem{Samba}
Hideo Sambe.
\newblock Steady states and quasienergies of a quantum-mechanical system in an
  oscillating field.
\newblock {\em Phys. Rev. A}, 7:2203, 1973.

\bibitem{Rigol_PRX14}
Luca D'Alessio and Marcos Rigol.
\newblock {Long-time Behavior of Isolated Periodically Driven Interacting
  Lattice Systems}.
\newblock {\em Phys. Rev. X}, 4:041048, 2014.

\bibitem{Ponte_AP15}
Pedro Ponte, Anushya Chandran, Z.~Papi{\'c}, and Dmitry~A. Abanin.
\newblock Periodically driven ergodic and many-body localized quantum systems.
\newblock {\em Annals of Physics}, 353:196, 2015.

\bibitem{srednicki_jpa99}
Mark Srednicki.
\newblock The approach to thermal equilibrium in quantized chaotic systems.
\newblock {\em Journal of Physics A}, 32:1463, 1999.

\bibitem{Polkovnikov_quergo:booklet}
Anatoli Polkovnikov.
\newblock Quantum ergodicity: fundamentals and applications, 2013.
\newblock Available at:
  {\texttt{http://physics.bu.edu/~asp/teaching/PY\_747.pdf}}.

\bibitem{Russomanno_EPL}
Angelo Russomanno, Rosario Fazio, and Giuseppe~E. Santoro.
\newblock Thermalization in a periodically driven fully connected quantum ising
  ferromagnet.
\newblock {\em Europhysics Letters}, 110:37005, 2015.

\bibitem{Sred_PRE94}
Mark Srednicki.
\newblock Chaos and quantum thermalization.
\newblock {\em Phys. Rev. E}, 50:888--901, 1994.

\bibitem{Deutsch_PRA91}
J.~M. Deutsch.
\newblock Quantum statistical mechanics in a closed system.
\newblock {\em Phys. Rev. A}, 43:2046--2049, 1991.

\bibitem{Rigol_Nat}
M.~Rigol, V.~Dunjko, and M.~Olshanii.
\newblock Thermalization and its mechanism for generic isolated quantum
  systems.
\newblock {\em Nature}, 452:854--858, 2008.

\bibitem{Bohigas_PRL84}
O.~Bohigas, M.~J. Giannoni, and C.~Schmit.
\newblock Characterization of chaotic quantum spectra and universality of level
  fluctuation laws.
\newblock {\em Phys. Rev. Lett.}, 52:1, 1984.

\bibitem{Berry_LH84}
M.~V. Berry.
\newblock {Semiclassical mechanics of regular and irregular motion}.
\newblock In R.~Stora G.~Ioos, R. H. G.~Helleman, editor, {\em {Les Houches,
  Session XXXVI, 1981 --- Chaotic Behaviour of Deterministic Systems}}, pages
  174--271. North-Holland Publishing Company, 1983.

\bibitem{Haake:book}
Fritz Haake.
\newblock {\em Quantum Signatures of Chaos ({$2^{\rm nd}$} ed)}.
\newblock Springer, 2001.

\bibitem{CHIRIKOV198877}
B.V. Chirikov, F.M. Izrailev, and D.L. Shepelyansky.
\newblock Quantum chaos: Localization vs. ergodicity.
\newblock {\em Physica D: Nonlinear Phenomena}, 33(1):77 -- 88, 1988.

\bibitem{Berry77}
M.V. Berry.
\newblock Regular and irregular semiclassical wavefunctions.
\newblock {\em Journal of Physics A}, 10:2083, 1977.

\bibitem{Pechukas}
P.~Pechukas.
\newblock Distribution of energy eigenvalues in the irregular spectrum.
\newblock {\em Phys. Rev. Lett.}, 51:943, 1983.

\bibitem{feingold_PRL}
Mario Feingold and Asher Peres.
\newblock Distribution of matrix elements of chaotic systems.
\newblock {\em Phys. Rev. A}, 34:591, 1986.

\bibitem{prosen_AP94}
T.~Prosen.
\newblock Statistical properties of matrix elements of Hamiltonian systems
  between integrability and chaos.
\newblock {\em Annals of Physics}, 235:115, 1994.

\bibitem{PhysRevE.50.888}
Mark Srednicki.
\newblock Chaos and quantum thermalization.
\newblock {\em Phys. Rev. E}, 50:888--901, Aug 1994.

\bibitem{Eckardt_PRE95}
Bruno Eckhardt, Shmuel Fishman, Jonathan Keating, Oded Agam, J\"org Main, and
  Kirsten M\"uller.
\newblock Approach to ergodicity in quantum wave functions.
\newblock {\em Phys. Rev. E}, 52:5893--5903, Dec 1995.

\bibitem{sred_therm}
M.~Srednicki.
\newblock Thermal fluctuations in quantized chaotic systems.
\newblock {\em Journal of Physics A}, 29(4):L75--L79, 1995.

\bibitem{Chirikov1979263}
Boris~V Chirikov.
\newblock A universal instability of many-dimensional oscillator systems.
\newblock {\em Physics Reports}, 52(5):263 -- 379, 1979.

\bibitem{chirikov}
G.~Casati, B.~V. Chirikov, J.~Ford, and Izrailev~F. M.
\newblock Stochastic behaviour of classical and quantum Hamiltonian systems.
\newblock In {\em Lecture Notes in Physics}, volume~93 of {\em TOGLIERE}, page
  334. Springer, 1979.

\bibitem{PhysRevLett.49.509}
Shmuel Fishman, D.~R. Grempel, and R.~E. Prange.
\newblock Chaos, quantum recurrences, and anderson localization.
\newblock {\em Phys. Rev. Lett.}, 49:509--512, Aug 1982.

\bibitem{PhysRevA.29.1639}
D.~R. Grempel, R.~E. Prange, and Shmuel Fishman.
\newblock Quantum dynamics of a nonintegrable system.
\newblock {\em Phys. Rev. A}, 29:1639, Apr 1984.

\bibitem{chirikov_argument}
Chirikov B.V., Izrailev F.M., and Shepelyansky D.L.
\newblock Dynamical stochasticity in classical and quantum mechanics.
\newblock {\em Sov. Sci. Rev.}, 2C(209), 1981.

\bibitem{PhysRev.109.1492}
P.~W. Anderson.
\newblock Absence of diffusion in certain random lattices.
\newblock {\em Phys. Rev.}, 109:1492--1505, Mar 1958.

\bibitem{Nekhoroshev1971}
N.~N. Nekhoroshev.
\newblock Behavior of Hamiltonian systems close to integrable.
\newblock {\em Functional Analysis and Its Applications}, 5(4):338--339, 1971.

\bibitem{PhysRevA.40.6130}
Kunihiko Kaneko and Tetsuro Konishi.
\newblock Diffusion in Hamiltonian dynamical systems with many degrees of
  freedom.
\newblock {\em Phys. Rev. A}, 40:6130, Nov 1989.

\bibitem{2018arXiv180101142R}
A.~{Rajak}, R.~{Citro}, and E.~G. {Dalla Torre}.
\newblock {Stability and pre-thermalization in chains of classical kicked
  rotors}.
\newblock {\em Journal of Physics A}, 51:465001, 2018.

\bibitem{nostrolavoro}
Simone Notarnicola, Fernando Iemini, Davide Rossini, Rosario Fazio, Alessandro
  Silva, and Angelo Russomanno.
\newblock From localization to anomalous diffusion in the dynamics of coupled
  kicked rotors.
\newblock {\em Phys. Rev. E}, 97:022202, Feb 2018.

\bibitem{2rot_1}
Borzumehr Toloui and Leslie~E. Ballentine.
\newblock Quantum localization for two coupled kicked rotors.
\newblock {\em arXiv:0903.4632v2 [quant-ph]}, Mar 2009.

\bibitem{2rot_2}
Fumihiro Matsui, Hiroaki~S. Yamada, and Kensuke~S. Ikeda.
\newblock Relation between irreversibility and entanglement in classically
  chaotic quantum kicked rotors.
\newblock {\em arXiv:1603.07050v1 [cond-mat.stat-mech]}, Mar 2016.

\bibitem{2rot_3}
Fumihiro Matsui, Hiroaki~S. Yamada, and Kensuke~S. Ikeda.
\newblock Lifetime of the arrow of time inherent in chaotic eigenstates: case
  of coupled kicked rotors.
\newblock {\em arXiv:1510.00199 [cond-mat.stat-mech]}, Oct 2015.

\bibitem{Adachi_PRL88}
S.~Adachi, M.~Toda, and K.~Ikeda.
\newblock Potential for mixing in quantum chaos.
\newblock {\em Phys. Rev. Lett.}, 61:659, 1988.

\bibitem{rozenbaum_e}
E.B. Rozenbaum and V.~Galitski.
\newblock Dynamical localization of coupled relativistic kicked rotors.
\newblock {\em Phys. Rev. B}, 95:064303, 2017.

\bibitem{rylands}
C.~Rylands, E.B. Rozenbaum, V.~Galitski, and R.~Konik.
\newblock Many-body localization in a kicked lieb-liniger gas.
\newblock {\em arXiv}, page 1904.09473, 2019.

\bibitem{shepelyanski_prl93}
D.L Shepelyansky.
\newblock Delocalization of quantum chaos by weak nonlinearity.
\newblock {\em Phys. Rev. Lett.}, 70:1787, 1993.

\bibitem{shepelyanski_prl08}
D.L Shepelyansky and A.~Pikovsky.
\newblock Destruction of anderson localization by a weak nonlinearity.
\newblock {\em Phys. Rev. Lett.}, 100:094101, 2008.
\sn{
\bibitem{Laptyeva_2014}
T V Laptyeva and M V Ivanchenko and S Flach.
\newblock Nonlinear lattice waves in heterogeneous media.
\newblock {\em Journal of Physics A: Mathematical and Theoretical}, 47, 2014.
\bibitem{PhysRevB.89.060301}
Ivanchenko, M. V. and Laptyeva, T. V. and Flach, S.
\newblock Quantum chaotic subdiffusion in random potentials.
\newblock {\em Phys. Rev. B}, 89, 2014.
}
\bibitem{marko}
Marko \ifmmode \check{Z}\else \v{Z}\fi{}nidari\ifmmode~\check{c}\else
  \v{c}\fi{}, Antonello Scardicchio, and Vipin~Kerala Varma.
\newblock Diffusive and subdiffusive spin transport in the ergodic phase of a
  many-body localizable system.
\newblock {\em Phys. Rev. Lett.}, 117:040601, Jul 2016.

\bibitem{Matthew}
M.~Rispoli, A.~Lukin, R~Shittko, M.~Eric~Tai, J.~L{\'e}onard, and M.~Greiner.
\newblock Quantum critical behavior at the many-body-localization transition.
\newblock {\em arXiv}, page 1812.06959, 2018.
\sn{
\bibitem{Qin2017}
P. Quin, A.  Andreanov, H. C. Park, and S. Flach.
\newblock Interacting ultracold atomic kicked rotors: loss of dynamical localization.
\newblock {\em Scientific Reports}, 41,  page 41139, 2017.
\bibitem{refId0}
A. Marino, G. Torre, and R. Citro.
\newblock Dynamical localization of interacting ultracold atomic kicked rotors.
\newblock {\em EPL}, 127, page 50008, 2019.
\bibitem{0295-5075-96-3-30004}
G. Gligorić, J. D. Bodyfelt, and S. Flach.
\newblock Interactions destroy dynamical localization with strong and weak chaos.
\newblock {\em EPL (Europhysics Letters)}, 96, page  30004, 2011.
\bibitem{Yusipov2017}
I. I. Yusipov, T. V. Laptyeva, A. Y. Pirova, I. B. Meyerov, S. Flach
and  M. V. Ivanchenko,
\newblock Quantum subdiffusion with two- and three-body interactions.
\newblock {\em The European Physical Journal B}, 90, page 66, 2017.
}





\bibitem{Luitz_PRL16}
David~J. Luitz and Yevgeny Bar~Lev.
\newblock Anomalous thermalization in ergodic systems.
\newblock {\em Phys. Rev. Lett.}, 117:170404, Oct 2016.

\bibitem{Roy_PRB18}
Sthitadhi Roy, Yevgeny~Bar Lev, and David~J. Luitz.
\newblock Anomalous thermalization and transport in disordered interacting
  floquet systems.
\newblock {\em Phys. Rev. B}, 98:060201, Aug 2018.

\bibitem{foini2019eigenstate}
Laura Foini and Jorge Kurchan.
\newblock Eigenstate thermalization and rotational invariance in ergodic
  quantum systems.
\newblock {\em arXiv preprint arXiv:1906.01522}, 2019.

\bibitem{Peres_PRA84}
Asher Peres.
\newblock Ergodicity and mixing in quantum theory. i.
\newblock {\em Phys. Rev. A}, 30:504, 1984.

\bibitem{oganesyan2007localization}
V.~Oganesyan and D.A. Huse.
\newblock {Localization of interacting fermions at high temperature}.
\newblock {\em Physical Review B}, 75(15):155111, 2007.

\bibitem{eynard}
B.~Eynard, T.~Kimura, and S.~Ribault.
\newblock Random matrices.
\newblock {\em arXiv}, page 1510.04430, 2015.

\bibitem{Berry375}
M.~V. Berry and M.~Tabor.
\newblock Level clustering in the regular spectrum.
\newblock {\em Proceedings of the Royal Society of London A: Mathematical,
  Physical and Engineering Sciences}, 356(1686):375--394, 1977.

\bibitem{Wegner1980}
F.~Wegner.
\newblock Inverse participation ratio in 2+$\epsilon$ dimensions.
\newblock {\em Zeitschrift f{\"u}r Physik B Condensed Matter}, 36(3):209--214,
  Sep 1980.

\bibitem{Edwards_JPC72}
J.~T. Edwards and D.~J. Thouless.
\newblock {Numerical studies of localization in disordered systems}.
\newblock {\em J. Phys. C}, 5:807, 1972.

\bibitem{PhysRevLett.79.1913}
V.~E. Kravtsov and K.~A. Muttalib.
\newblock New class of random matrix ensembles with multifractal eigenvectors.
\newblock {\em Phys. Rev. Lett.}, 79:1913--1916, Sep 1997.

\bibitem{0305-4470-19-8-004}
C~Castellani and L~Peliti.
\newblock Multifractal wavefunction at the localisation threshold.
\newblock {\em Journal of Physics A: Mathematical and General}, 19(8):L429,
  1986.

\bibitem{PhysRevB.84.134209}
Alberto Rodriguez, Louella~J. Vasquez, Keith Slevin, and Rudolf~A. R\"omer.
\newblock Multifractal finite-size scaling and universality at the anderson
  transition.
\newblock {\em Phys. Rev. B}, 84:134209, Oct 2011.

\bibitem{PhysRevB.96.104201}
Maksym Serbyn, Z.~Papi\ifmmode~\acute{c}\else \'{c}\fi{}, and Dmitry~A. Abanin.
\newblock Thouless energy and multifractality across the many-body localization
  transition.
\newblock {\em Phys. Rev. B}, 96:104201, Sep 2017.

\bibitem{chalker1996spectral}
JT~Chalker, VE~Kravtsov, and IV~Lerner.
\newblock Spectral rigidity and eigenfunction correlations at the anderson
  transition.
\newblock {\em Journal of Experimental and Theoretical Physics Letters},
  64(5):386--392, 1996.

\bibitem{PhysRevLett.83.61}
JT~Chalker, VE~Kravtsov, and IV~Lerner.
\newblock Measurement-Induced Quantum Diffusion.
\newblock {\em Phys. Rev. Lett.},
  84:61--64, Jul 1999.
  
\bibitem{d2001quantum}
M~d'Arcy, R~Godun, M~Oberthaler, D~Cassettari, and G~Summy.
  Quantum enhancement of momentum diffusion in the delta-kicked rotor. {\em Phys. Rev. Lett.}, {87:074102}, Jul. 2001.  

\bibitem{ammann1998quantum}
H~Ammann, R~Gray, I~Shvarchuck, and N~Christensen. 
\newblock{Quantum   delta-kicked rotor: Experimental observation of decoherence.} 
\newblock {\em Phys. Rev. Lett.}, 80:4111--4115, May 1998

\bibitem{Molinari:banded}
Luca Molinari.
\newblock Band random matrices, kicked rotator and disordered systems.
\newblock In Zbigniew Haba, Wojciech Cegla, and Lech Jak{\'o}bczyk, editors,
  {\em Stochasticity and Quantum Chaos}, volume 317 of {\em Mathematics and Its
  Applications}. SPRINGER-SCIENCE+BUSINESS MEDIA, B.V., 1995.

\bibitem{mario_jpa97}
Mario Feingold.
\newblock Localization in strongly chaotic systems.
\newblock {\em Journal of Physics A}, 30:3603–3612, 1997.

\end{thebibliography}

\providecommand{\noopsort}[1]{}\providecommand{\singleletter}[1]{#1}%


%
%
%

\end{document}